\documentclass[sigconf, nonacm]{acmart}

\usepackage{amsmath,amsfonts}
\usepackage{algorithmic}
\usepackage{graphicx}
\usepackage{textcomp}
\usepackage{xcolor}
\usepackage{caption}
\usepackage{subcaption}
\usepackage{multirow}
\usepackage{listings}
\usepackage{comment}

\newcommand{\subfigsizea}{0.48\textwidth}

\newtheorem{obs}{\textbf{Observation}}

\newtheorem{qst}{\textbf{Question}}

\lstdefinestyle{customc}{
belowcaptionskip=1\baselineskip,
breaklines=true,
frame=L,
xleftmargin=\parindent,
language=C,
showstringspaces=false,
basicstyle=\footnotesize\ttfamily,
keywordstyle=\bfseries\color{green!40!black},
commentstyle=\itshape\color{purple!40!black},
identifierstyle=\color{blue},
stringstyle=\color{orange},
morekeywords={uint32_t},
}

\def\BibTeX{{\rm B\kern-.05em{\sc i\kern-.025em b}\kern-.08em
    T\kern-.1667em\lower.7ex\hbox{E}\kern-.125emX}}
\begin{document}

\title{Blockchain Goes Green? \\ Part II: Characterizing the Performance and
Cost of Blockchains on the Cloud and at the Edge}

\author{Dumitrel Loghin}
\affiliation{%
  \institution{School of Computing \\ National University of Singapore}
}
\email{dumitrel@comp.nus.edu.sg}

\author{Tien Tuan Anh Dinh}
\affiliation{%
  \institution{Information Systems Technology and Design \\
Singapore University of Technology and Design}
}
\email{dinhtta@comp.nus.edu.sg}

\author{Aung Maw}
\affiliation{%
  \institution{Information Systems Technology and Design \\
Singapore University of Technology and Design}
}
\email{aung\_maw@comp.nus.edu.sg}

\author{Chen Gang}
\affiliation{%
  \institution{College of Computer Science and Technology, Zhejiang University, China}
}
\email{cg@zju.edu.cn}

\author{Yong Meng Teo}
\affiliation{%
  \institution{School of Computing \\ National University of Singapore}
}
\email{teoym@comp.nus.edu.sg}

\author{Beng Chin Ooi}
\affiliation{%
  \institution{School of Computing \\ National University of Singapore}
}
\email{ooibc@comp.nus.edu.sg}

\begin{abstract}

While state-of-the-art permissioned blockchains can achieve thousands of
transactions per second on commodity hardware with x86/64 architecture, their
performance when running on different architectures is not clear. The goal of
this work is to characterize the performance and cost of permissioned
blockchains on different hardware systems, which is important as diverse
application domains are adopting t. To this end, we conduct extensive cost and
performance evaluation of two permissioned blockchains, namely Hyperledger
Fabric and ConsenSys Quorum, on five different types of hardware covering both
x86/64 and ARM architecture, as well as, both cloud and edge computing. The
hardware nodes include servers with Intel Xeon CPU, servers with ARM-based
Amazon Graviton CPU, and edge devices with ARM-based CPU. Our results reveal a
diverse profile of the two blockchains across different settings, demonstrating
the impact of hardware choices on the overall performance and cost. We find that
Graviton servers outperform Xeon servers in many settings, due to their powerful
CPU and high memory bandwidth. Edge devices with ARM architecture, on the other
hand, exhibit low performance. When comparing the cloud with the edge, we show
that the cost of the latter is much smaller in the long run if manpower cost is
not considered.

\end{abstract}

\maketitle

\thispagestyle{plain}
\pagestyle{plain}

\section{Introduction}

Blockchain is increasingly adopted by users and businesses around the globe.
While blockchain technology became famous through \textit{public},
\textit{permissionless} chains, such as Bitcoin~\cite{Bitcoin_2008} and
Ethereum~\cite{Ethereum_2013} which are used primarily for cryptocurrency
transfers, there is a growing interest for \textbf{permissioned
blockchains}\footnote{In this paper, we only analyze \textit{permissioned}
blockchains. Unless otherwise specified, the term blockchain in this paper means
permissioned blockchain.}. These permissioned blockchains, also termed as
\textit{private}, \textit{consortium} or \textit{enterprise} blockchains, allow
only authenticated entities to be part of the system. As such, they often relax
the security aspect of the consensus protocol by employing Byzantine
fault-tolerant (BFT) alternatives to Proof-of-Work (PoW) such as Practical
Byzantine Fault Tolerance (PBFT)~\cite{PBFT_99} or even crash fault-tolerant
(CFT) protocols such as Raft~\cite{raft_14}. State-of-the-art permissioned
blockchains can achieve thousands of transactions per second, by improving
consensus, network performance, or by adopting database
techniques~\cite{Hung_SIGMOD_19,HotStuff_PODC_19,ResilientDB_VLDB_20}. However,
we note that these systems are designed for traditional enterprise workloads,
and are evaluated only on commodity or cluster-grade servers with x86/64
architecture, being it at the edge or on the cloud. For example,
AHL~\cite{Hung_SIGMOD_19} uses general instances on Google Cloud Platform
spanning multiple regions.

As blockchains are maturing, we observe that more and more applications are
adopting them. Besides the traditional enterprise data processing
applications, such as payment systems, emerging blockchain applications include
blockchain cloud services~\cite{qldb, chainstack}, and edge
computing~\cite{fabric_iot, helium}.
These applications come with different hardware resource demands, that is, they
may not run efficiently on x86/64 commodity hardware~\cite{Dumi_BlockchainGreen,
s1, s2}. Besides x86/64, we have witnessed the proliferation of ARM
architecture, which first started with mobile devices. Recently, ARM has been
adopted as an alternative architecture in cloud computing~\cite{graviton}.
However, the performance and cost of running blockchains on ARM-based hardware
are not clear. As a result, users cannot make an informed decision of what
system to run their blockchain on.

Our goal is to provide an in-depth analysis of blockchain performance and cost
on different CPU architectures, which helps inform the design of emerging
blockchain applications. To this end, we select five representative edge and
cloud systems, including servers with Intel Xeon CPU, ARM-based Amazon Graviton
instances~\cite{graviton}, high-end and low-end edge devices with ARM CPU
represented by Nvidia Jetson TX2~\cite{Jetson_TX2_17} and Raspberry Pi
4~\cite{Pi4_Specs}, respectively.
We run two representative permissioned blockchains on these hardware nodes,
namely, Hyperledger Fabric~\cite{Hyperledger_2018, fabric_code} and ConsenSys
Quorum~\cite{quorum_code}. The former is widely used in the industry and well
studied by the database community in the last few years~\cite{Dinh_SIGMOD_2017,
Hung_SIGMOD_19, RPC_SIGMOD_20, RPC_SIGMOD_21, Sharma_SIGMOD_19}. Fabric adopts
an execute-order-validate transaction flow and currently uses Raft as a
consensus mechanism for its ordering service. Quorum~\cite{quorum_code} adopts a
more traditional order-execute transaction flow. We use throughput and latency
as performance metrics when running the two blockchains on the selected
hardware. Our cost metric consists of both the fixed hardware cost and the
continuing energy cost. We note that our work is the first to  quantify the cost
of operating a permissioned blockchain.

In summary, the following key contributions are presented in this paper:
\begin{itemize}
  
  \item We conduct a systematic performance analysis of two representative
  permissioned blockchains, namely, Fabric and Quorum, on five types of systems
  covering both x86/64 and ARM architectures, as well as both edge and cloud setups.
  These five types of systems are (i) AWS cloud instances with Intel Xeon CPU,
  (ii) AWS cloud instances with ARM Graviton CPU, (iii) edge-based servers with
  Intel Xeon CPU, (iv) edge-based Nvidia Jetson TX2 with ARM CPU, and (v)
  edge-based Raspberry Pi 4 with ARM CPU.
  
  \item We show that ARM-based Graviton instances are more cost-effective than
  Intel Xeon instances. For example, Graviton achieves close to 10\% higher
  throughput for Fabric and 25\% lower throughput for Quorum compared to Xeon,
  while being 35\% cheaper. This, in turn, stems from the impressive CPU,
  memory, and networking sub-systems of Graviton. For example, we show that a
  Graviton instance exhibits $2\times$ higher main memory bandwidth compared to
  the Xeon-based instance.
  
  \item On the other hand, we show that both Fabric and Quorum do not
  efficiently utilize the available hardware resources. For example, the average
  number of cores used by Fabric and Quorum is 1.3 and 0.6, respectively. In
  addition, Fabric exhibits a high volume of cache misses that hinders
  efficient execution.
  
  \item Finally, we show that hosting blockchain nodes at the edge (on-premise)
  may be cheaper if the manpower cost for operating and maintaining them is
  disregarded. In contrast, adding manpower cost leads to higher cost at
  the edge compared to the cloud.

\end{itemize}

The rest of this paper is organized as follows. In Section~\ref{sec:rel_work} we
give an overview of blockchain frameworks and present related works that analyze
the performance of blockchains. In Section~\ref{sec:setup} we describe our
experimental setup, and in Section~\ref{sec:analysis} we conduct our in-depth
analysis. We conclude the paper in Section~\ref{sec:concl}.

\section{Background and Related Work}
\label{sec:rel_work}

In this section, we provide a background on blockchain and survey the related
work on time, energy, and cost analysis of blockchains. Due to space
constraints, our background on blockchain is brief. We direct the reader to
other works~\cite{Dinh_TKDE_2018, RPC_SIGMOD_21, RPC_Blockchain_2022} for more
details on blockchain systems.

\subsection{Blockchain}

A blockchain is a distributed ledger managed by a network of mutually
distrusting nodes (or peers). The ledger is stored as a linked list (or chain)
of blocks, where each block consists of transactions. The links in the chain are
built using cryptographic pointers to ensure that no one can tamper with the
chain or with the data inside a block.

Blockchains are most famous for being the underlying technology of
cryptocurrencies, but many can support general-purpose applications. This
ability is determined by the execution engine and data model. For example,
Bitcoin~\cite{Bitcoin_2008} supports only operations related to cryptocurrency
(or token) manipulation. On the other hand, Ethereum~\cite{Ethereum_2013} can
run arbitrary computations on its Turing-complete Ethereum Virtual Machine
(EVM). At the data model level, there are at least three alternatives used in
practice. The \textit{Unspent Transaction Output} (\textit{UTXO}) model, used by
Bitcoin among others, represents the ledger states as transaction ids and
associated unspent amounts that are the input of future transactions. The
\textit{account/balance} model resembles a classic banking ledger. A more
generic model used by Hyperledger Fabric consists of \textit{key-value states}.
On top of the data model, developers can write general applications that operate
on the blockchain's states. Such applications are called~\textit{smart
contracts}. In this paper, we extend the smart contracts from
Blockbench~\cite{Dinh_SIGMOD_2017} to support Fabric (v2.3.1) and Quorum.

Depending on how nodes can join the network, the blockchain is \textit{public}
(or \textit{permissionless}) or \textit{private} (or \textit{permissioned}). In
public networks, anybody can join or leave and, thus, the security risks are
high. Most of the cryptocurrency blockchains are public, such as
Bitcoin~\cite{Bitcoin_2008} and Ethereum~\cite{Ethereum_2013}. On the other
hand, private blockchains allow only authenticated peers to join the network.
Typically, private blockchains, such as Fabric~\cite{Hyperledger_2018} and
Quorum~\cite{quorum_code}, are deployed inside or across big organizations.

Blockchains operate in a network of mutually distrusting peers, where some peers
may not be just faulty but malicious. Hence, they assume a Byzantine
environment~\cite{Lamport_Byzantine}, in contrast to the crash-failure model
used by the majority of distributed systems. For example, Proof-of-Work (PoW) is
a BFT consensus mechanism where participating nodes, called \textit{miners},
need to solve a difficult cryptographic puzzle. The miner that solves the puzzle
first has the right to append transactions to the ledger. Since this mining
process is both time and energy inefficient, some alternatives have been
proposed, such as Proof-of-Stake (PoS)~\cite{eth_pos}, Proof-of-Authority (PoA),
and Proof-of-Elapsed-Time (PoET)~\cite{poet_sawtooth}. In PoS, the nodes need to
set aside some coins (the \textit{stake}) based on which the validator of each
block is chosen. In case a malicious validator is detected, it loses its stake.
Through this mechanism, malicious behavior is discouraged. In PoA, some nodes
with known identity, called \textit{validators}, are trusted to validate all the
transactions. In case a trusted node acts maliciously, its reputation is
affected and it may be removed from the network. In PoET, each node needs to
wait for a random period, before being able to propose a new block. The node
with the smallest wait period is the one appending the next block. 

On the other hand, PBFT~\cite{PBFT_99} consists of exchanging $O(n^2)$ messages
among the nodes to reach an agreement on the transactions to be appended to the
chain. However, BFT consensus does not scale well and leads to low blockchain
throughput~\cite{Dinh_SIGMOD_2017, RPC_SIGMOD_21}. This is one reason why
permissioned blockchains started to replace BFT with CFT consensus. For example,
both Fabric and Quorum support Raft consensus~\cite{raft_14}. Another reason for
using CFT consensus in permissioned chains is the higher accountability due to
node authentication in such platforms.

\subsection{Performance Analysis of Blockchains}

There are a number of related works that analyze the performance of
blockchains~\cite{Dinh_TKDE_2018}, \cite{Dinh_SIGMOD_2017},
\cite{Pongnumkul_17}, \cite{Thakkar_mascots_18}, \cite{XU2021102436},
\cite{RPC_SIGMOD_21}. However, only a few include energy or cost
analysis~\cite{ala2014bitcoin}, \cite{Dumi_BlockchainGreen},
\cite{Sankaran_ICDCS_2018}, \cite{Suankaewmanee_ICNC_18}, but their analysis is
of limited depth.
 
Sankaran et al.~\cite{Sankaran_ICDCS_2018} analyze the time and energy
performance of an in-house Ethereum network consisting of high-performance
mining servers and low-power Raspberry Pi 3 clients. These low-power nodes
cannot run Ethereum mining due to their limited memory size, hence, they only
take the role of clients. In this paper, we focus on permissioned blockchains
and use Quorum as a representative of Ethereum-based blockchains. On the other
hand, we use low-power devices with higher performance, such as Jetson TX2 and
Raspberry Pi 4.
 
MobiChain~\cite{Suankaewmanee_ICNC_18} is an approach that allows mining on
mobile devices running Android OS, in the context of mobile e-commerce. While it
provides an analysis of both time and energy performance, MobiChain has no
comparison to other blockchains. In terms of energy analysis, the authors show
that it is more energy-efficient to group multiple transactions in a single
block since there is less mining work and therefore less time and power wasted
in this process. However, larger blocks increase latency and result in a poor
user experience.

Jupiter~\cite{Jupiter_ICDE_18} is a blockchain designed for mobile devices. It
aims to address the problem of storing a large ledger on mobile devices with
limited storage capacity. The testbed in~\cite{hl_rpi_2019} is based on 14
Raspberry Pi 3 nodes running Hyperledger Fabric version 1.0. However, there is
no time or energy performance evaluation in both of these
works~\cite{Jupiter_ICDE_18},~\cite{hl_rpi_2019}.

Ruan et al.~\cite{RPC_SIGMOD_21} compare blockchains with distributed database
systems using a taxonomy with four dimensions, namely, replication, concurrency,
storage, and sharding. Similar to this paper, they analyze Fabric and Quorum but
their analysis is only focusing on the time performance and it is conducted on
commodity x86/64 servers. Complementary to our analysis of blockchains on
different hardware systems, \cite{RPC_SIGMOD_21} evaluates the effect of
different blockchain and benchmarking parameters, such as record size, block
size, replication model, failure model, among others.

Blockbench~\cite{Dinh_SIGMOD_2017} is a benchmarking suite comprising both
simple (micro) benchmarks and complex (macro) benchmarks. The microbenchmarks,
namely \textit{CPUHeavy}, \textit{IOHeavy}, and \textit{Analytics} are stressing
different hardware subsystems such as the CPU, memory, and IO. At the same time,
the microbenchmarks evaluate the performance of different blockchain layers. For
example, \textit{CPUHeavy} evaluates the performance of the execution engine,
while \textit{IOHeavy} evaluates the performance of the data storage. The macro
benchmarks are represented by \textit{YCSB}, \textit{Smallbank}, and
\textit{Donothing}. The \textit{YCSB} macro benchmark implements a key-value
storage, while \textit{Smallbank} represents OLTP and simulates basic banking
operations. The \textit{Donothing} benchmark is used to evaluate the consensus
protocol since it does not engage the execution and data storage layers. The
performance in terms of throughput and latency is evaluated on traditional
high-performance servers with Intel Xeon CPU. We note that Blockbench analyzes
the performance of Fabric v0.6 with PBFT. In this paper, we extend Blockbench to
support Fabric v2.3.1 and Quorum v20.10.0. Moreover, we do not analyze only the
time performance, but also the power and cost of a wider range of node types,
including both x86/64 and ARM architectures.
To the best of our knowledge, we provide the first extensive time, energy, and
cost performance analysis of permissioned blockchains on both x86/64 and ARM
architectures, as well as both at the edge and on the cloud.

\section{Experimental Setup}
\label{sec:setup}

In this section, we describe our experimental setup in terms of blockchains,
benchmarks, and hardware nodes.

\begin{figure}[tp]
	\centering    
	\includegraphics[width=0.47\textwidth]{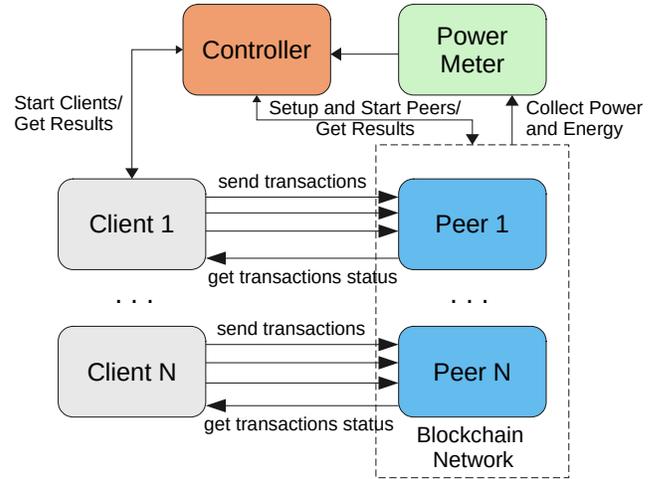}
	\caption{Experimental Approach}
    \label{fig:approach}    
\end{figure}

\subsection{Overview}

We illustrate our experimental approach in Figure~\ref{fig:approach}. The setup
consists of a blockchain network comprising $N$ peers, a set of clients that
send transactions to the peers, a controller, and a power meter to collect power
and energy measurements. For simplicity, we consider $N$ clients, each client
sending transactions to one blockchain peer. In reality, a client can send
transactions to multiple peers. The controller node is used to set up the
benchmarking environment, start the peers, set up the smart contracts, start the
clients, and collect the results, including power and energy.

\subsection{Blockchains and Benchmarks}

The permissioned blockchain frameworks analyzed in this paper are Hyperledger
\textbf{Fabric} (v2.3.1) and ConsenSys \textbf{Quorum} (v20.10.0).
Fabric~\cite{fabric_code} is a crash fault-tolerant (CFT) permissioned
blockchain that adopts an execute-order-validate transaction flow. In a Fabric
network, there are three types of nodes: peers, orderers, and clients. A client
sends a transaction request to a set of peers, depending on an endorsement
policy. For example, an \textit{AND} policy including all the peers means that
the client needs to send the transaction and receive endorsements from all the
peers. A peer processes the transaction request and creates read and write sets
to mark which states are touched by the transaction. However, the peer does not
persist the transaction's effects on its local database. When the client gets
all the endorsements from the peers, it sends the transaction to the orderers
such that they will pack it in a block. Once a block is formed, the orderers
send it to all the peers. Lastly, each peer validates the transactions in a
block and persists the changes to its local database.
Fabric v2.3.1 uses Raft~\cite{raft_14} as the consensus mechanism among the
orderers, and LevelDB~\cite{Leveldb_2011} as its local database.

Quorum~\cite{quorum_code} is a permissioned blockchain that started as a fork of
Ethereum. Hence, Quorum supports Solidity smart contracts, but it uses different
consensus mechanisms than Ethereum because of its permissioned nature.
In this paper, we analyze both a CFT Quorum that uses Raft as the consensus
protocol among the peers, as well as a BFT version that uses Istanbul BFT
(IBFT)~\cite{moniz2020istanbul} as the consensus protocol. In contrast to
Fabric, a Quorum network has only peers and clients, and the transaction flow is
order-execute. This means that transactions are first grouped into blocks and
then executed by each peer in the network. Similar to Fabric, Quorum uses
LevelDB as its local database.

For benchmarking these blockchains, we extend Blockbench\footnote{The Blockbench
source code used in our experiments can be found at
\url{https://github.com/dloghin/blockbench/tree/analysis2021}.}~\cite{Dinh_SIGMOD_2017}
to support Fabric v2.3.1 and Quorum v20.10.0. That is, we implemented Fabric and
Quorum smart contracts for the key-value store benchmark and the scripts needed
to run the benchmark. We use the YCSB macro-benchmark consisting of 50\% read
and 50\% write operations of single key-value pairs of 1kB in size. One or more
client nodes send transactions to the blockchain peers in an asynchronous mode.
That is, the request threads of the clients do not wait for the result of a
transaction. Instead, a status thread periodically queries the blockchain to get
the committed transactions.

In our experiments, the block size of Fabric is set to 500 transactions and the
runtime for the benchmark is set to 120s. In Quorum, there is no limit to block
size. The runtime is set to at least 240s to account for a slower startup
compared to Fabric. We vary the number of blockchain peers from four to ten and
we use the client request rate that leads to the best performance. In this
paper, we report the best results out of three runs for each experiment. We note
that the standard deviation is lower than 10\% of the mean for each experiment.

\subsection{Hardware Systems}

\label{sec:sys_specs}

In this paper, we are analyzing the performance of five types of systems
(hardware nodes) covering both edge and cloud computing, as well as both x86/64
and ARM architectures. The specifications of these systems are summarized in
Table~\ref{table:sys_char}. We measure the power and energy of the edge nodes
with a Yokogawa power meter connected to the AC lines. We report the AC power
values in this paper.

At the edge, we measure the time performance and power of the following three
types of nodes. First, we have x86/64 nodes with Intel Xeon E5-1650 v3 CPU
clocked at 3.5~GHz, 32~GB DDR3 memory, 2~TB hard-disk (HDD), and 1~Gbps
networking interface card (NIC). These nodes, termed \textbf{Xeon(edge)} in this
paper, run Ubuntu 18.04. Second, we use high-end ARM-based devices represented
by Nvidia Jetson \textbf{TX2}~\cite{Jetson_TX2_17}. A TX2 node has a
heterogeneous 6-core 64-bit CPU with two NVIDIA Denver cores and four ARM
Cortex-A57 cores clocked at more than 2GHz. Each node has 8~GB LPDDR4, a 32~GB
SD card, and 1~Gbps NIC. TX2 is equipped with an integrated low-power GPU.
However, we are not using the GPU in our experiments. The TX2 nodes are running
Ubuntu 16.04 which is officially supported by Nvidia on these systems. Third, we
use low-end ARM-based devices represented by Raspberry Pi 4 model B
(\textbf{RP4})~\cite{Pi4_Specs}.
An RP4 has a 4-core ARM Cortex-A72 CPU of 64-bit ARM architecture, 8~GB of
low-power DDR4 memory, a 64~GB SD card that acts as storage, and 1~Gbps NIC. RP4
runs the beta version of the 64-bit Debian-based Raspberry Pi
OS~\cite{rpios_64bit}.

On the cloud, we use two types of instances from Amazon Web Services (AWS) to
represent both x86/64 and ARM architectures. AWS was chosen because it is the
only widely-known cloud provider that offers ARM-based instances. We use
\textit{m5n.2xlarge}~\cite{m5n_Specs} AWS instances to represent the x86/64
architecture. At the time of running the experiments, one such instance costs
\$0.476 per hour in the US West (Oregon) region. These nodes, termed
\textbf{Xeon(cloud)} in this paper, are equipped with Intel Xeon Platinum 8259CL
CPUs clocked at 2.5~GHz. Each node has 8 CPU cores, 32~GB RAM, 50~GB SSD
storage, and up to 25~Gbps networking. A Xeon(cloud) node runs Ubuntu 18.04 OS.
Then, we use \textit{ large}~\cite{m6g_Specs} AWS instances to represent the
emerging 64-bit ARM server market. These instances, termed \textbf{Graviton} in
this paper, are based on Amazon's Graviton processors that use 64-bit ARM
Neoverse architecture~\cite{graviton}. Each node has 8 CPU cores, 32~GB RAM,
50~GB SSD, up to 10~Gbps networking, and runs Ubuntu 18.04. At the time of
running the experiments, one such instance costs \$0.308 per hour in the US West
(Oregon), being $1.55\times$ or 35\% cheaper than one Xeon(cloud)
\textit{m5n.2xlarge} instance.

In summary, we note that cloud nodes use CPUs with more cores, higher frequency,
bigger cache size, and more levels of cache, as shown in Table 1. Moreover,
cloud nodes have bigger RAM sizes and faster networking. We shall see in the
next sections how these hardware characteristics impact the performance of
blockchain systems.

\begin{figure*}[tp]
	\centering
    \begin{subfigure}{0.47\textwidth}
	\includegraphics[width=0.99\textwidth]{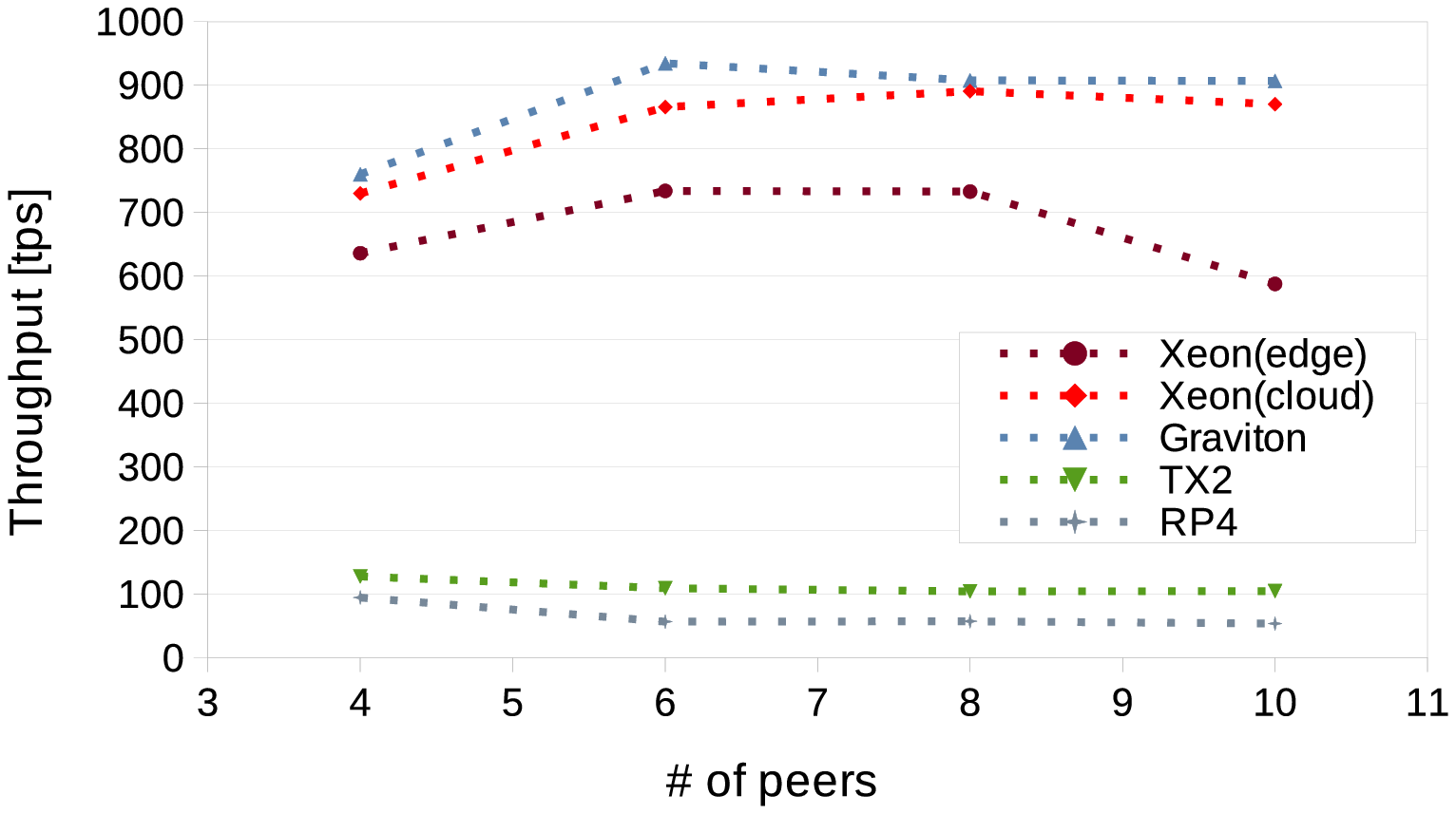}
	\caption{Throughput of Fabric.}
    \label{fig:fabric-tps}
    \end{subfigure}
    ~
    \centering
    \begin{subfigure}{0.47\textwidth}
	\includegraphics[width=0.99\textwidth]{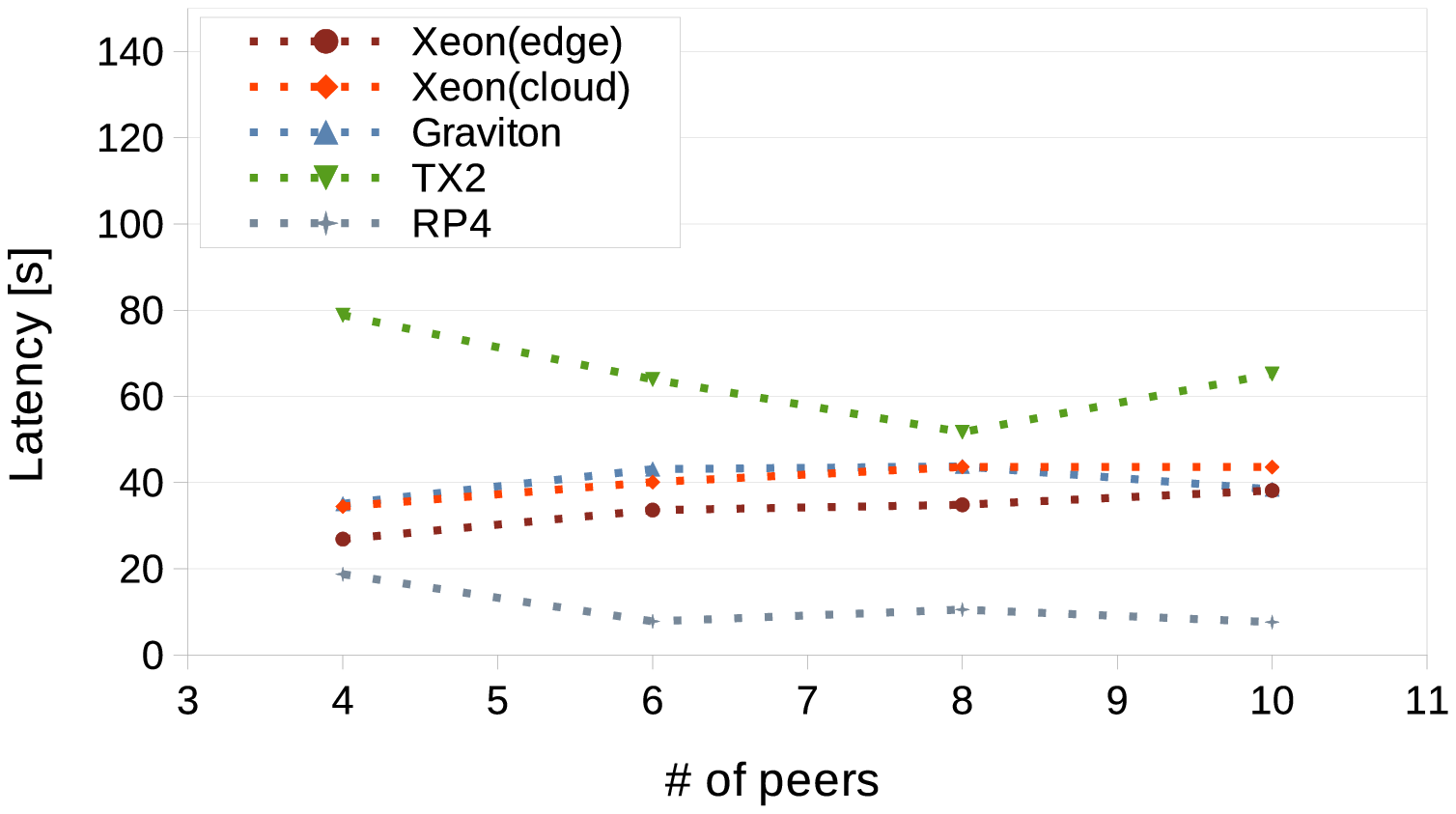}
	\caption{Latency of Fabric.}
    \label{fig:fabric-latency}
    \end{subfigure}
        \quad
    \centering
    \begin{subfigure}{0.47\textwidth}
	\includegraphics[width=0.99\textwidth]{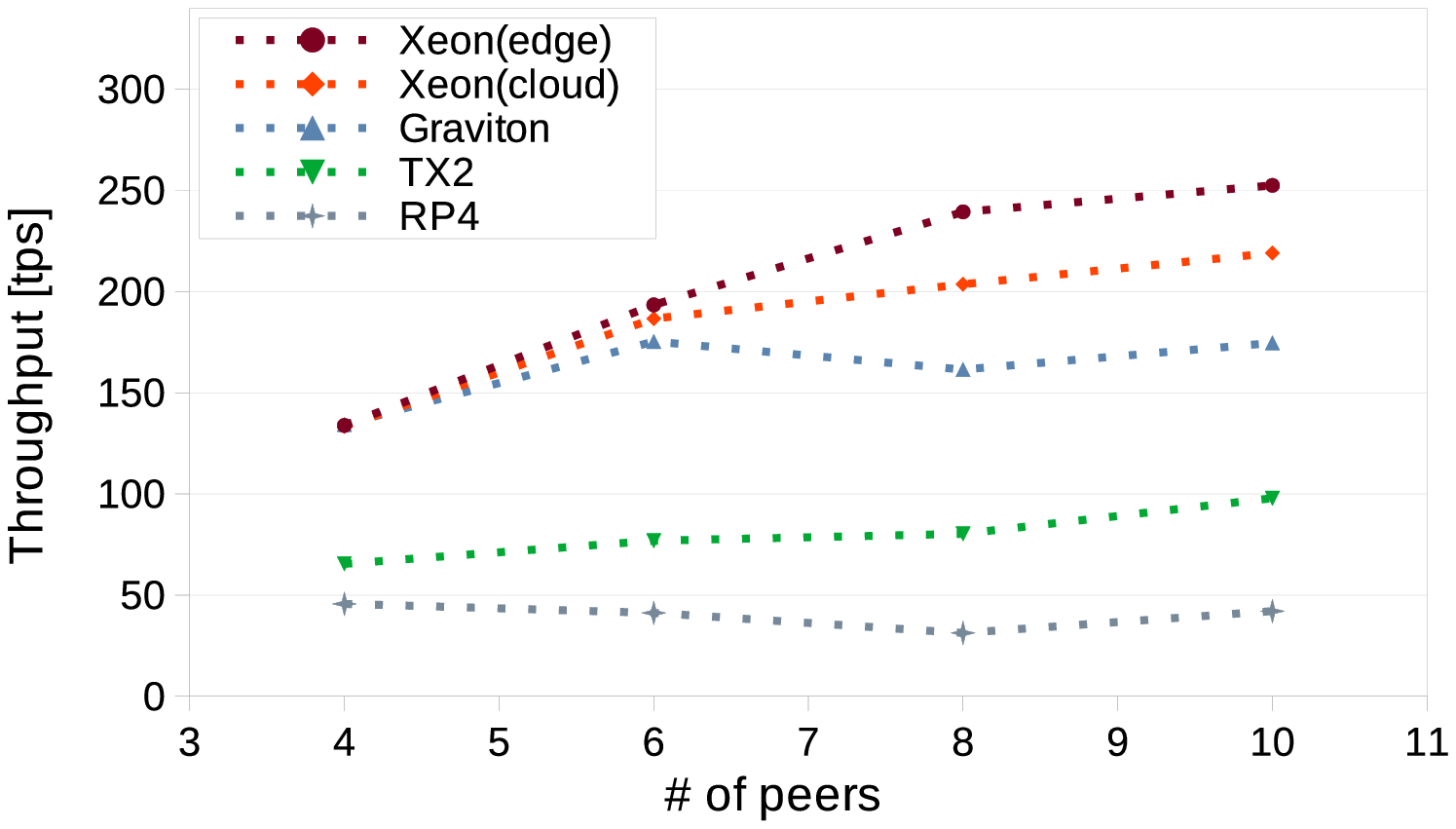}
	\caption{Throughput of Quorum(Raft).}
    \label{fig:quorum-raft-tps}
    \end{subfigure}
    ~
    \centering
    \begin{subfigure}{0.47\textwidth}
	\includegraphics[width=0.99\textwidth]{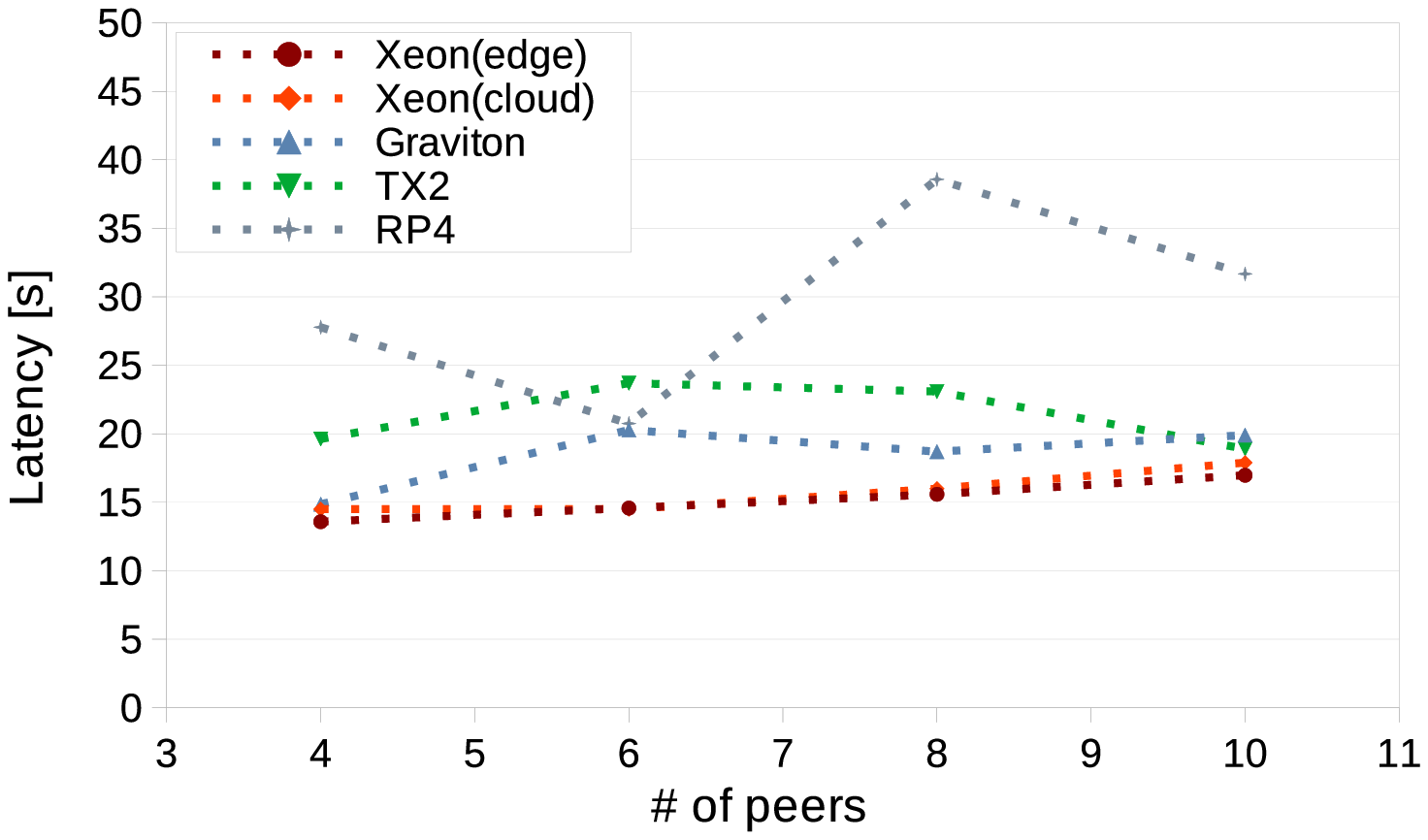}
	\caption{Latency of Quorum(Raft).}
    \label{fig:quorum-raft-latency}
    \end{subfigure}
         \quad
    \centering
    \begin{subfigure}{0.47\textwidth}
	\includegraphics[width=0.99\textwidth]{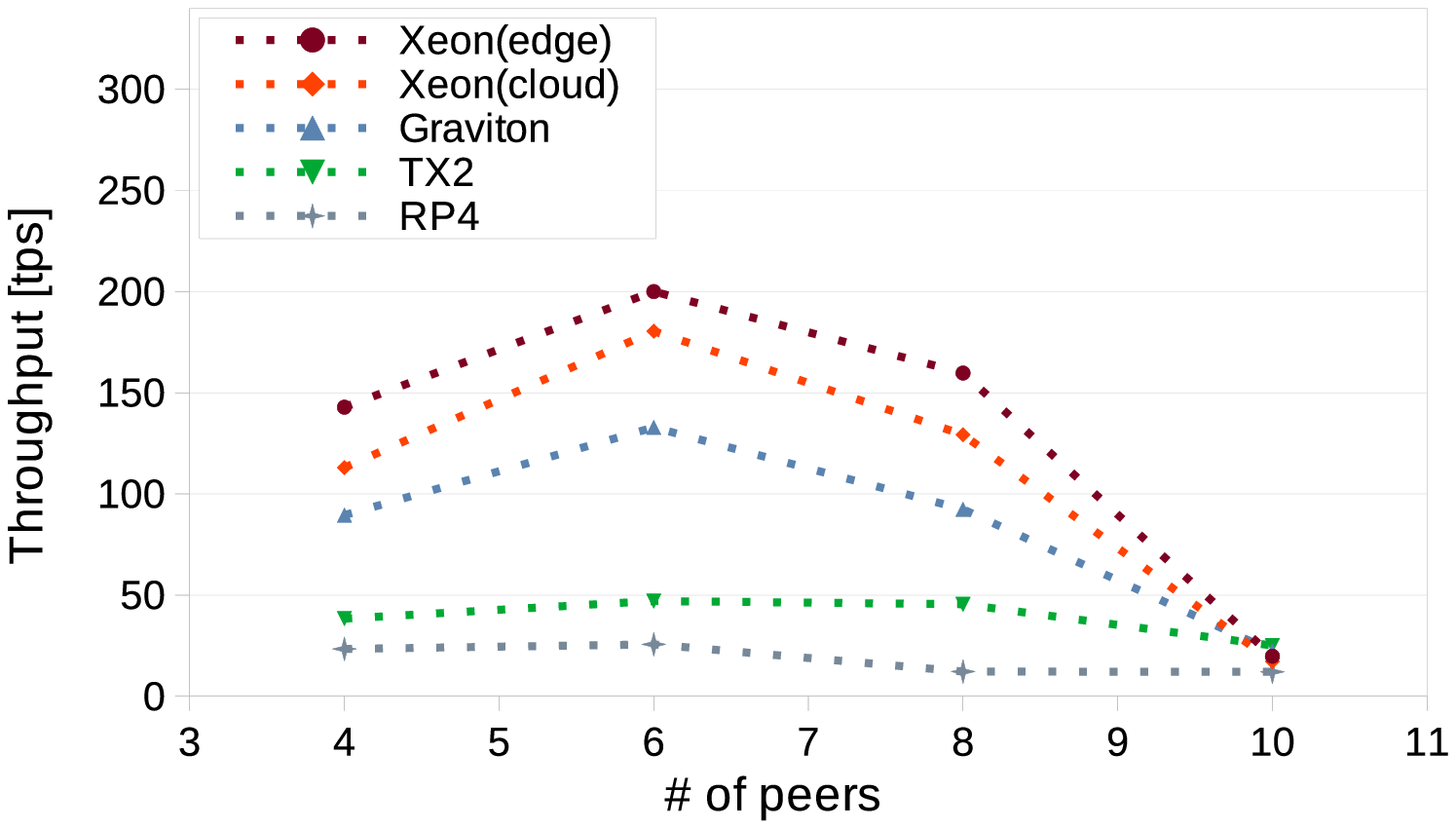}
	\caption{Throughput of Quorum(IBFT).}
    \label{fig:quorum-ibft-tps}
    \end{subfigure}
    ~
    \centering
    \begin{subfigure}{0.47\textwidth}
	\includegraphics[width=0.99\textwidth]{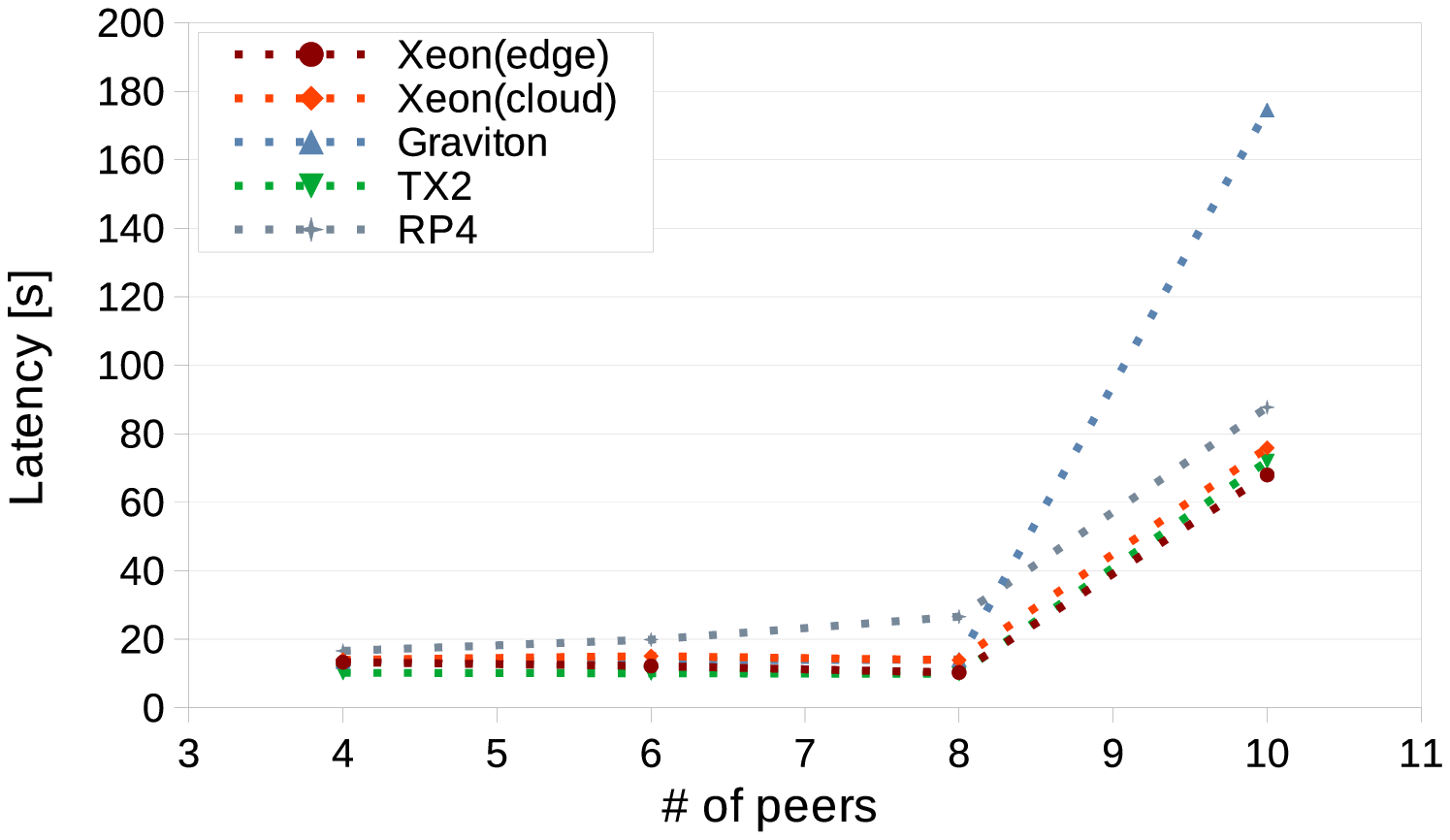}
	\caption{Latency of Quorum(IBFT).}
    \label{fig:quorum-ibft-latency}
    \end{subfigure}
        \quad
    \centering
    \begin{subfigure}{0.47\textwidth}
	\includegraphics[width=0.99\textwidth]{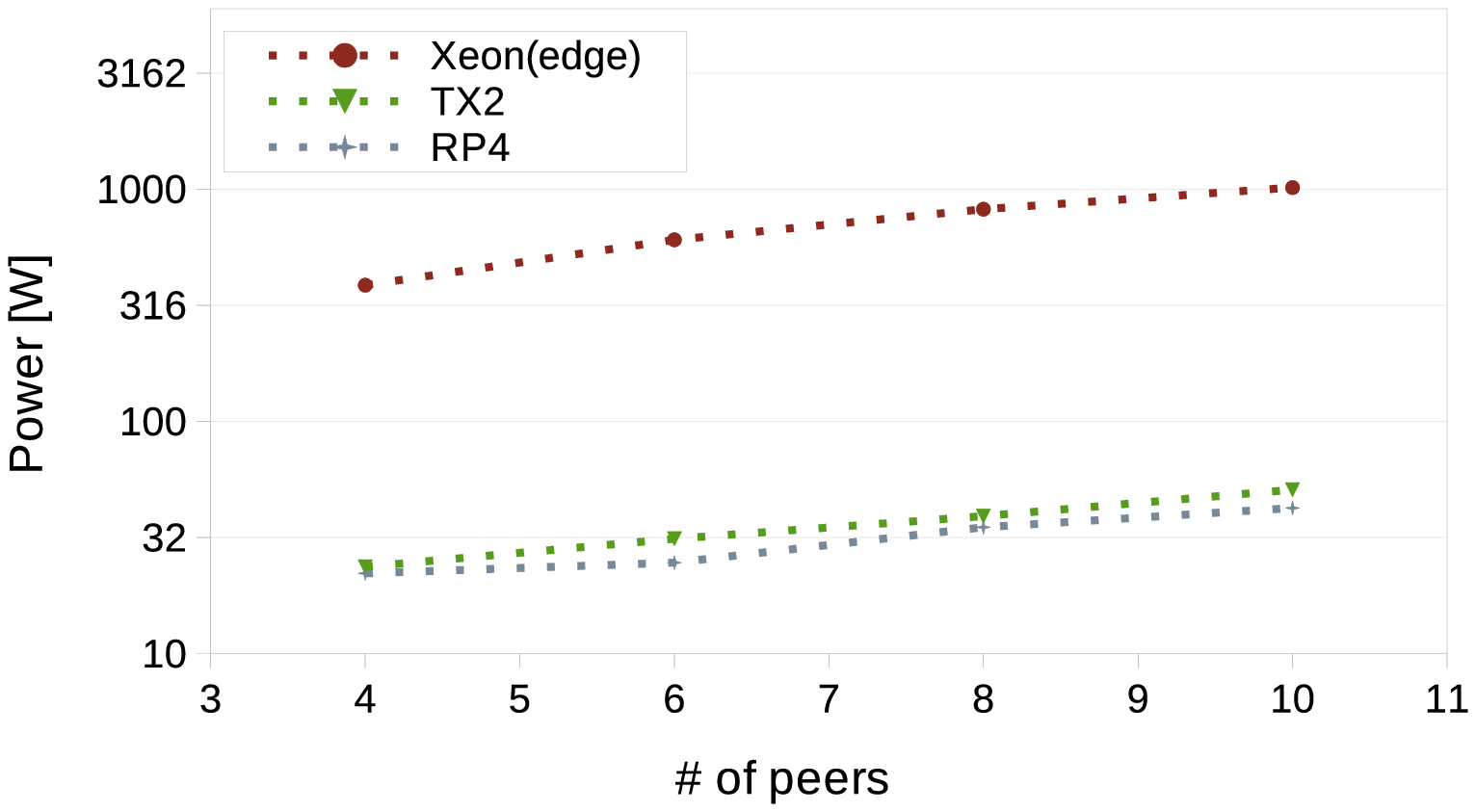}
	\caption{Power usage of Fabric.}
    \label{fig:fabric-power}
    \end{subfigure}
     ~
    \centering
    \begin{subfigure}{0.47\textwidth}
	\includegraphics[width=0.99\textwidth]{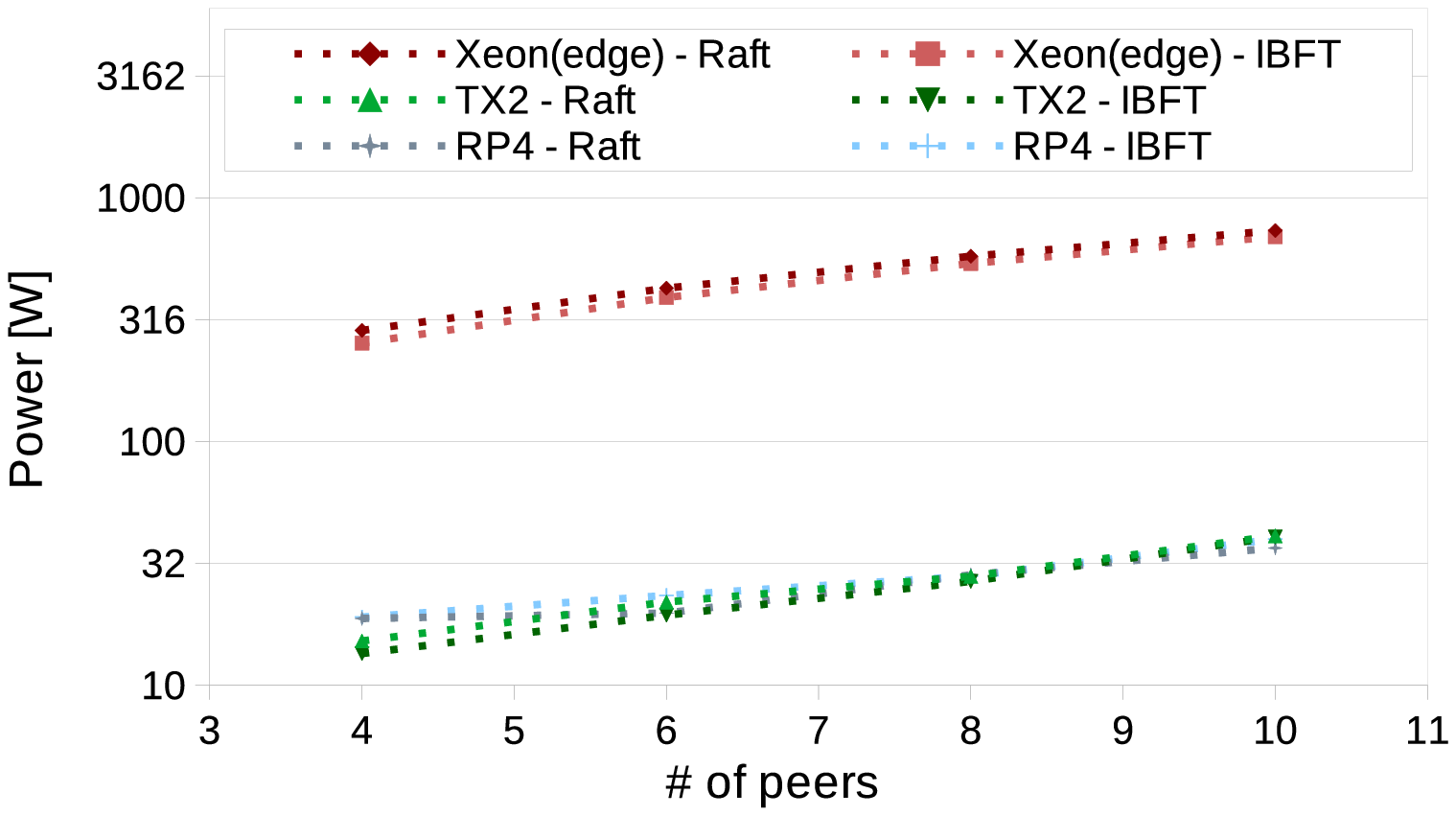}
	\caption{Power usage of Quorum.}
    \label{fig:quorum-power}
    \end{subfigure}
    
	\caption{Throughput, latency, and power with increasing number of peers.}
	\label{fig:perf-vs-peers}
\end{figure*}

\section{Performance Analysis}
\label{sec:analysis}

In this section, we conduct a systematic performance analysis of the selected
blockchains and hardware nodes.

\begin{table*}[t]
\centering
\caption{Hardware systems characterization.}
\label{table:sys_char}
\begin{tabular}{|l|l|r|r|r|r|r|}
\hline
& \multicolumn{1}{c|}{\multirow{2}{*}{\textbf{Characteristic}}} &
\multicolumn{5}{c|}{\textbf{System}} \\
\cline {3-7} & & \bf{Xeon(edge)} & \bf{Xeon(cloud)} & \bf{Graviton} & \bf{TX2} & \bf{RP4} \\
\hline
\hline
\multirow{9}{*}{Specs} & ISA & x86-64 & x86-64 & AARCH64 & AARCH64 & AARCH64/ARMv7l \\
& Cores & 6 (12) & 8 & 8 & 6 & 4 \\
& Frequency & 1.2-3.5~GHz & 2.5~GHz & 2.5~GHz & 0.346-2.04~GHz &
0.6-1.5~GHz \\
& L1 Data Cache & 32~kB & 32~kB & 64~kB & 32-128~kB & 32~kB\\
& L2 Cache & 256~kB (core) & 1~MB (core) & 1~MB (core) & 2~MB & 1~MB\\
& L3 Cache & 12~MB & 35.8~MB & 32~MB & N/A & N/A \\ 
& Memory & 32~GB DDR3 & 32~GB DDR4 & 32~GB & 8~GB LPDDR4 & 8~GB LPDDR4 \\
& Storage & 1~TB SSD & 50~GB SSD & 50~GB SSD & 64~GB SD card & 64~GB SD card \\
& Networking & Gbit & up to 25~Gbit & up to 10~Gbit & Gbit & Gbit \\
\hline
\multirow{5}{*}{CPU} & CoreMark (one core) [IPS] & 25,201.6 & 24,061.6 &
20,266.2 & 9,936.1 & 8,555.1 \\
& System power [W] & 70.6 & - & - & 4.2 & 3.6 \\
\cline{2-7}
& CoreMark (all cores) [IPS] & 170,864.7 & 137,126.8 & 162,054.4 & 68,092.3 &
34,255.1 \\
& System power [W] & 115.5 & - & - & 10.4 & 5.7 \\
\cline{2-7}
& Idle system power [W] & 50.8 & - & - & 2.4 & 1.7 \\
\hline
\multirow{5}{*}{Storage} & Write throughput [MB/s] & 160 & 132 & 58.3 & 16.3 &
19.0 \\
& Read throughput [MB/s] & 172 & 134 & 141 & 89 & 46 \\
& Buffered read throughput [GB/s] & 8.1 & 6.5 & 6.9 & 2.7 & 1.3 \\
& Write latency [ms] & 9.3 & 0.51 & 0.49 & 17.1 & 1.33 \\
& Read latency [ms] & 2.5 & 0.22 & 0.22 & 2.8 & 0.72 \\
\hline
\multirow{3}{*}{Network} & TCP bandwidth [Mbits/s] & 941 & 9530 & 9680 & 943 &
943 \\
& UDP bandwidth [Mbits/s] & 810 & 4540 & 5800 & 546 & 957 \\
& Ping latency [ms] & 0.14 & 0.1 & 0.09 & 0.3 & 0.15 \\
\hline
\end{tabular}
\end{table*}

\subsection{Overview}

We first highlight the key findings of our measurement-based evaluation,
followed by a systematic in-depth analysis of the hardware systems and
blockchain frameworks. These key observations are based on the throughput,
latency, and power results presented in Figure~\ref{fig:perf-vs-peers} for all
three blockchains on all five types of nodes.

\begin{obs}
\label{obs:o1}
Surprisingly, Graviton exhibits up to 8\% higher performance compared 
to Xeon(cloud) when running Fabric, as shown in Figure~\ref{fig:fabric-tps}. 
On the other hand, Graviton exhibits 20\% and 26\% lower throughput when
running Quorum(Raft) and Quorum(IBFT), respectively, compared to Xeon(cloud), as shown in 
Figure~\ref{fig:quorum-raft-tps} and Figure~\ref{fig:quorum-ibft-tps}, respectively. 
In the context of 35\% lower cost of Graviton compared to Xeon(cloud),
the ARM-based server is more cost-efficient.
\end{obs}

\begin{obs}
\label{obs:o2}
As shown in Figure~\ref{fig:perf-vs-peers}, Xeon(cloud) exhibits up to 50\%
higher throughput compared to Xeon(edge) when running Fabric, while Xeon(edge)
exhibits up to 26\% higher throughput compared to Xeon(cloud) when running Quorum.
\end{obs}

\begin{obs}
\label{obs:o3}
Surprisingly, the performance gap between Graviton and
TX2 (or RP4) is big when running Fabric. As shown in
Figure~\ref{fig:fabric-tps}, TX2 exhibits a throughput that is $6-9\times$ lower
compared to Graviton. On the other hand, the throughput of TX2 running Quorum is
only $2-3\times$ lower compared to Graviton.
\end{obs}

\begin{obs}
\label{obs:o4}
While we expected TX2 to exhibit higher performance than RP4, it is interesting
to observe that TX2 is also more power-efficient. This is because TX2 offers
more performance per unit of energy compared to RP4, even if the latter uses
less power.
\end{obs}

\begin{obs}
\label{obs:o5}
The cost of hosting blockchain nodes at the edge is much lower compared to the
cloud, in the long run, when manpower cost associated with operating and
maintaining the edge nodes is disregarded. Conversely, adding the
manpower cost leads to almost double cost at the edge compared to the cloud.
\end{obs}

These observations give rise to a series of questions for which we seek answers
in the following sections. Here, we outline some of these questions.
\begin{qst}
\label{qst:q1}
Why does Graviton achieve higher performance than Xeon(cloud) when running
Fabric, while Xeon(cloud) has higher performance when running Quorum?
\end{qst} 

\begin{qst}
\label{qst:q2}
Why does Xeon(cloud) achieve higher performance than Xeon(edge) when running
Fabric, while Xeon(edge) has higher performance when running Quorum? 
\end{qst}
 
\begin{qst}
\label{qst:q3}
Why is there such a big performance gap between TX2 and Graviton when
running Fabric?
\end{qst}

\begin{qst}
\label{qst:q4}
Why is the power efficiency of RP4 lower compared to TX2?
\end{qst}

\begin{qst}
\label{qst:q5}
Where and under what performance-cost circumstances should blockchain nodes be
hosted: at the edge or on the cloud?
\end{qst}

\subsection{Systems Characterization}
\label{sec:sys_char}

Before answering the above questions related to the blockchain frameworks, we
characterize the hardware systems using a series of benchmarks that stress key
sub-systems, such as CPU, memory, storage, and networking. 

\begin{figure*}[!t]
\centering
\begin{subfigure}{\subfigsizea}
\centering
\includegraphics[width=\textwidth]{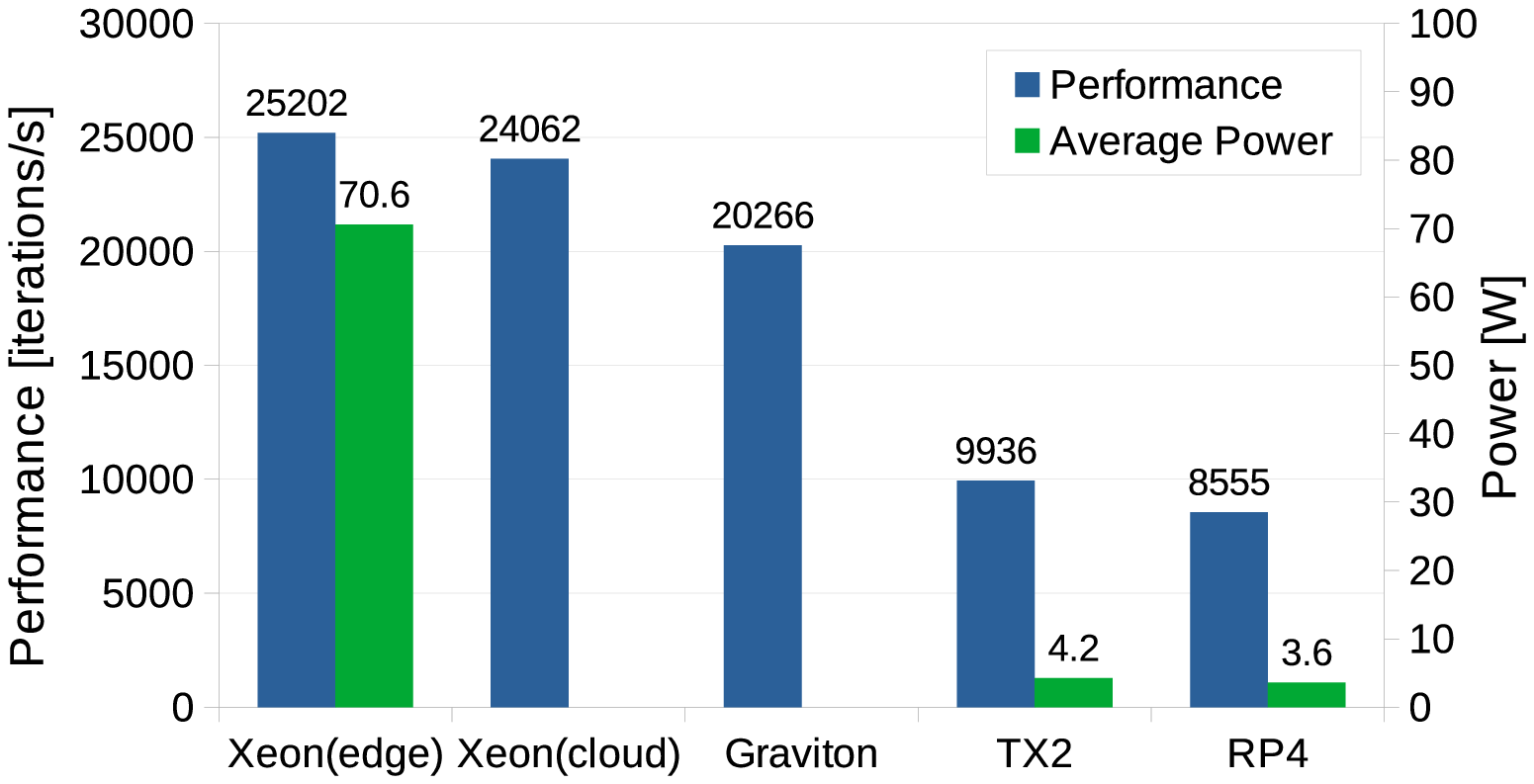}
\caption{CoreMark on one core}
\label{fig:coremark_one_core}
\end{subfigure}
~~~
\begin{subfigure}{\subfigsizea}
\centering
\includegraphics[width=\textwidth]{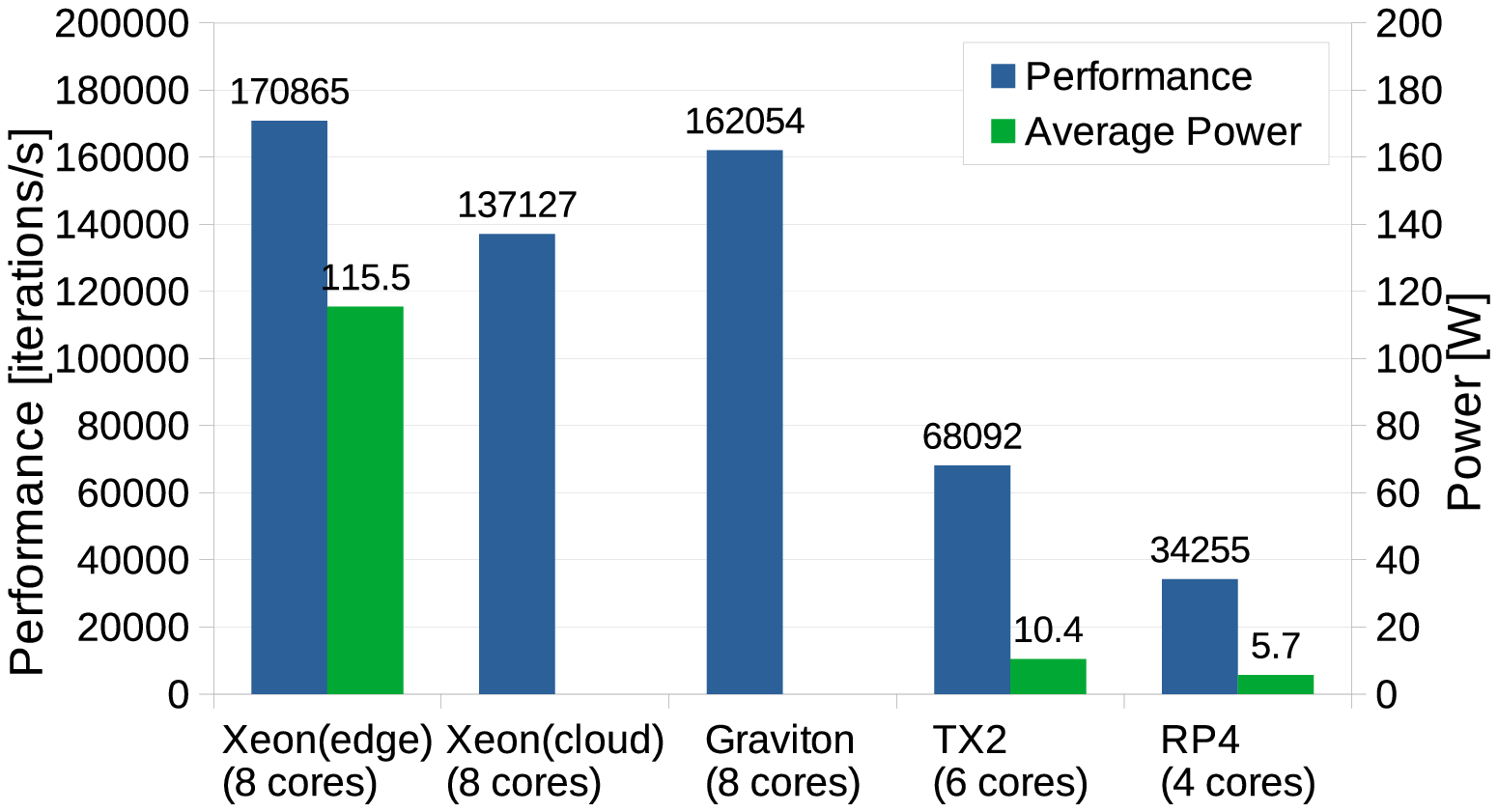}
\caption{CoreMark on all cores}
\label{fig:coremark_all_cores}
\end{subfigure}
\caption{Performance and power at CPU level.}
\label{fig:sys_cpu}
\end{figure*}

To assess the performance of the CPU of each type of node, we first use
CoreMark~\cite{ARM_CoreMark_09}, a modern benchmark from Embedded Microprocessor
Benchmark Consortium (EEBMC) designed to characterize CPU cores of both x86/64
and ARM architectures. CoreMark estimates the CPU performance in terms of
iterations per second (IPS). In Figure~\ref{fig:coremark_one_core} and
Figure~\ref{fig:coremark_all_cores}, we show the performance and average power
usage of the five systems running CoreMark on a single core and all cores,
respectively. For multi-core analysis, we enable all available cores, including
virtual cores in systems that support Hyper-threading. The exception is
Xeon(edge) which has 12 virtual cores, but we benchmark only 8 of them for a
fair comparison with Xeon(cloud) and Graviton.

At the single-core level, Xeon(edge) achieves the highest performance, followed
by Xeon(cloud), Graviton, TX2, and RP4, which are 1.05, 1.2, 2.5, and 2.9 times
slower, respectively. Even if Xeon(edge) has a newer Xeon CPU, it exhibits lower
performance compared to Xeon(edge) due to lower clock frequency. That is, the
Intel Xeon E5-1650 of Xeon(edge) runs at 3.5~GHz, while the Intel Xeon Platinum
8259CL of Xeon(cloud) runs at 2.5~GHz. Nevertheless, the newer Intel Xeon
Platinum 8259CL is more efficient since it yields higher performance per GHz. On
the other hand, Graviton achieves impressive performance for an ARM-based
CPU~\cite{Dumi_PARCO_2019}. Even if its clock frequency is not officially
stated, our measurements suggest that Graviton CPU runs at 2.5~GHz. Lastly, the
performance of TX2 and RP4 is lower compared to the other systems, but their
power consumption is also much lower. For example, TX2 and RP4 use $17\times$
and $20\times$, respectively, less power compared to Xeon(edge).

At the multi-core level, the surprise is Graviton which achieves higher
performance than Xeon(cloud) on 8 cores, as shown in
Figure~\ref{fig:coremark_all_cores}. Even if the Intel Xeon Platinum 8259CL CPU
has 24 physical cores, our measurements suggest that a \textit{m5n.2xlarge}
instance uses only 4 physical cores, while the other 4 are Hyper-threading
cores.
In contrast, a \textit{m6g.2xlarge} instance uses 8 physical Graviton cores.
This leads to a higher performance of Graviton with 8 cores compared to
Xeon(cloud). On the other hand, RP4 suffers from having only 4 cores, being
$5\times$ and $2\times$ slower than Xeon(edge) and TX2, respectively.

\begin{figure*}[!t]
\centering
\begin{subfigure}{\subfigsizea}
\centering
\includegraphics[width=0.96\textwidth]{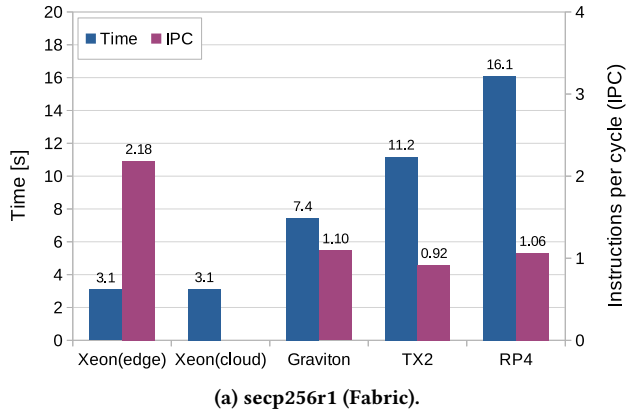}
\caption{secp256r1 (Fabric).}
\label{fig:ecdsa-sign}
\end{subfigure}
~~~
\begin{subfigure}{\subfigsizea}
\centering
\includegraphics[width=\textwidth]{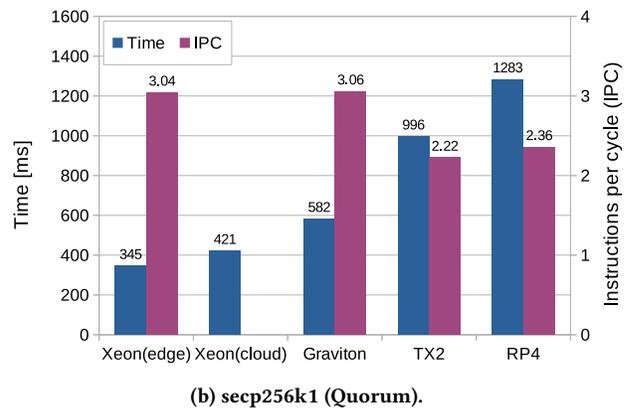}
\caption{secp256k1 (Quorum).}
\label{fig:secp256-sign}
\end{subfigure}
\caption{Performance of ECDSA signing operations}
\label{fig:ecdsa}
\end{figure*}

\begin{figure}[tp]
\centering
\includegraphics[width=0.48\textwidth]{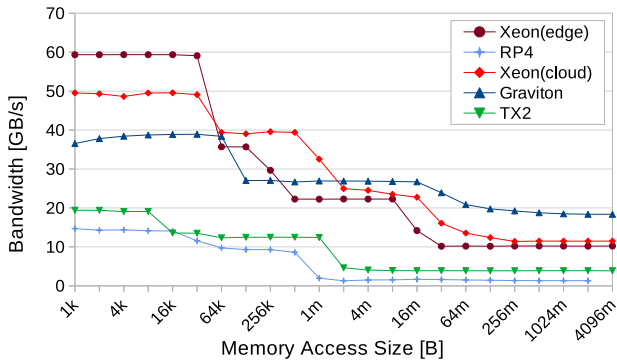}
\caption{Memory bandwidth.}
\label{fig:sys_mem}
\end{figure}

A significant aspect of blockchain is the usage of cryptography operations which
are, in general, compute-intensive. Both Fabric and Quorum  use elliptic curve,
namely ECDSA with \textit{secp256r1} and with \textit{secp256k1} respectively.
Our second CPU benchmarking assesses the performance of such cryptography
operations. Our profiling of Fabric shows that a significant proportion of the
CPU time is spent in \texttt{p256MulInternal} and \texttt{p256SqrInternal}
functions (see Listing~\ref{lst:perf_fabric}) which are part of the
\texttt{crypto/ecdsa} Go package. This package implements the elliptic curve
cryptography using \textit{secp256r1}, also known as \textit{NIST
P-256}~\cite{IETF_EC}. We isolated the ECDSA operations in Fabric's Go code and
measure the performance of both sign and verify operations. For this
benchmarking, we run 100,000 operations on messages of 64~B in size. The results
for signing, shown in Figure~\ref{fig:ecdsa}, follow the same trend as the
CoreMark benchmark. The verification has similar trends, just that it takes
around $3\times$ longer than signing on all the systems. We also observe that
Xeon systems are more efficient as they can pack around two instructions per
cycle (IPC) compared to around one instruction per cycle for the ARM-based
systems. We note that performance counters could not be obtained for Xeon(cloud)
due to AWS restrictions.

In Quorum, a significant part of the CPU time is spent in the
\texttt{secp256k1\_fe\_sqr\_inner} and \texttt{secp256k1\_fe\_mul\_inner} \quad
calls (see Listing~\ref{lst:perf_quorum}) which are part of the
\texttt{crypto/secp256k1} package. \textit{Secp256k1} is a way to compute the
elliptic curve which is also employed in Bitcoin~\cite{Secp256k1}. Hence, we
benchmark the sign operation which uses this function and we show the results in
Figure~\ref{fig:secp256-sign}. Interestingly, this cryptographic algorithm is
around 22\% slower on Xeon(cloud) compared to Xeon(edge). We attribute this
difference to the higher clock frequency of Xeon(edge) and the fact that
\textit{secp256k1} is very compute-intensive and highly optimized. This can be
observed from the high IPC yielded on all the systems. On Xeon and Graviton,
more than three instructions are executed in one cycle, while on TX2 and RP4
there are more than two instructions per cycle. For such an application, the CPU
clock frequency matters more, hence, Xeon(edge) is the fastest system.

Next, we assess the performance of the memory sub-system in terms of read-write
bandwidth measured with \textit{lmbench}~\cite{lmbench}.
The results in Figure~\ref{fig:sys_mem} represent the read-write bandwidth
measured with \textit{lmbench}~\cite{lmbench}.
At level one cache (L1), Xeon systems have the highest bandwidth, and this can
be correlated with the clock frequency of the cores. However, at the main memory
level, we have a surprise:
Graviton exhibits close to 20~GB/s, two times more than the Xeon systems. This
is a remarkable feature of the system designed by Amazon. TX2 and RP4 exhibit
main memory bandwidths of around 4~GB/s and 1.3~GB/s, respectively. This low
bandwidth, together with the relatively small memory size hinder the execution
of modern workloads on these ARM-based edge nodes.

The storage IO sub-system plays a key role in blockchain since the ledger and
the state database need to be written to persistent storage. Hence, we measure
the bandwidth and latency of the storage sub-system using \textit{dd} and
\textit{ioping} Linux commands, respectively. The results expose a complex
landscape. While the Xeon systems exhibit high bandwidth for both reads and
writes, the ARM-based systems expose asymmetric performance where the write
operations are a few times slower than the read operations. This is expected
since TX2 and RP4 are equipped with SD cards and Graviton is equipped with
NVMe-based SSD. On the other hand, the latency of both reads and writes is
higher on Xeon(edge) compared to Xeon(cloud) and we attribute this to mechanical
hard-disk (HDD) versus solid-state disk (SSD).

At the networking level, we measure the bandwidth and latency using
\textit{iperf} and \textit{ping} Linux commands, respectively, and summarize the
results in Table~\ref{table:sys_char}. As per their specifications, the
cloud-based systems have higher bandwidth which is close to 10~Gbps. In
contrast, edge-based systems have bandwidths close to 1~Gbps. The slightly
higher \textit{ping} latency of TX2 and RP4 can be attributed to the lower clock
frequency of these wimpy nodes. To validate this hypothesis, we measured the
networking latency while disabling the DVFS by fixing the clock frequency. TX2
supports twelve frequency steps in the range 346~MHz-2.04~GHz. RP4 supports ten
frequency steps in the range 600~MHz-1.5~GHz. We obtained Pearson correlation
coefficients of -0.93 and -0.84 for TX2 and RP4, respectively, between the
frequency and networking latency. These coefficients suggest strong inverse
proportionality and expose the impact of CPU processing on the networking stack.

Lastly, we discuss the difference in using the experimental 64-bit OS on RP4
compared to the official 32-bit OS. A 64-bit OS better matches the 64-bit ARM
CPU of RP4, and this is well highlighted by the benchmarking results. For
example, there is a 10\% improvement in CoreMark performance when using the
64-bit OS, while consuming the same power. For ECDSA the difference is much more
significant. Both the signing and verification are $8\times$ faster on the
64-bit OS. This is due to the very inefficient implementation of the algorithms
on the 32-bit Instruction Set Architecture (ISA). On 32-bit, there are $8\times$
more instructions executed compared to the 64-bit ISA. At memory and networking
levels we did not observe any significant difference between the two types of
OS. However, we observed a slight improvement in storage performance. There is a
$2\times$ improvement in both direct write and buffered read bandwidths. This is
due to more efficient drivers for the direct write, and double access size for
the buffered read which uses the main memory. In summary, we recommend the use
of the 64-bit OS on RP4 for improved performance.

\subsection{Performance Analysis}
\label{sec:bcperf}

In this section, we aim to answer the questions related to the performance of
the blockchain frameworks. We start by answering the first part of
\textbf{Question~\ref{qst:q1}}, that is, why does Graviton achieve higher
performance than Xeon(cloud) when running Fabric? Based on our benchmarking in
Section~\ref{sec:sys_char}, the advantage of Graviton may stem from its higher
performance when using all the cores or from its higher memory bandwidth
compared to Xeon(cloud). To further investigate this, we profile Fabric with
\textit{perf} Linux tool. We observe that the average number of CPU cores used
at runtime is below 1.3, meaning that Fabric mostly uses a single core. Hence,
Graviton could not take advantage of its higher multi-core performance. On the
other hand, memory references have a higher impact on the performance of Fabric.
Our analysis shows between 5 and 6.5 billion last-level cache (LLC) references
in 120 seconds, out of which 10\% are misses. Let us suppose that all LLC
references are hits, each reference takes 60 cycles~\cite{haswell}, the
references are not overlapped, and the clock frequency is 3.5 GHz. This results
in 90-110s out of the 120s spent in accessing the LLC. Although this is a
simplified analysis, it shows that Fabric's execution is cache/memory-intensive.
As such, Graviton achieves higher performance due to its higher LLC and main
memory bandwidth.

In contrast, the execution of Quorum exhibits around 4.5 and 7.6 billion LLC
references in 360s on follower peers and the leader peer, respectively. With the
same assumptions as above, the time spent in accessing the LLC is around 80s, or
131s for the leader, out of the 360s of runtime. Hence, Quorum is less
cache/memory-intensive compared to Fabric. Even at CPU core utilization, Quorum
is less intensive compared to Fabric since it uses only 0.6 and 0.8 cores, on
average, with Raft and IBFT, respectively. Hence, Quorum exhibits higher
performance on the Xeon-based systems compared to Graviton, but this is because
of poor utilization of the hardware resources.

Next, we answer \textbf{Question~\ref{qst:q2}}, namely, why does Xeon(cloud)
achieve higher performance than Xeon(edge) when running Fabric, while Xeon(edge)
has higher performance when running Quorum? We show in
Section~\ref{sec:sys_char} that Xeon(edge) has better CPU performance but lower
storage access latency performance and networking bandwidth compared to
Xeon(cloud). Hence, we investigate the effect of slower storage and different
networking bandwidth on Fabric. To investigate the effect of slower storage, we
introduce artificial delays during the read and write operations in the YCSB
smart contract. However, we do not observe any difference in Fabric's
throughput.

\begin{figure}[tbp]
\centering
\includegraphics[width=0.48\textwidth]{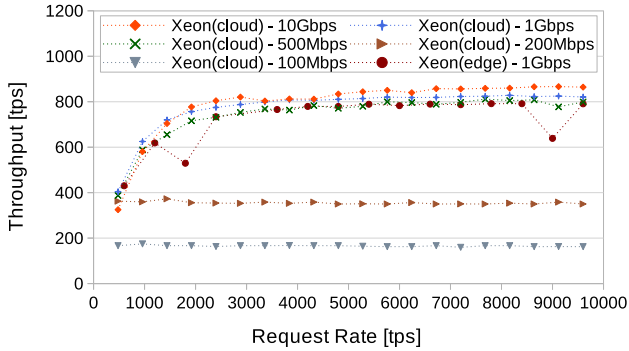}
\caption{Effect of networking bandwidth on Fabric.}
\label{fig:fabric-networking}
\end{figure}

To investigate the effect of slower networking, we use \textit{tc} Linux tool to
limit the bandwidth of the networking interface on each Xeon(cloud) peer of the
Fabric network. As shown in Figure~\ref{fig:fabric-networking} for six Fabric
peers and one orderer, networking bandwidth has a significant impact on the
performance of Fabric. When limiting the bandwidth of Xeon(cloud) to 1~Gbps or
500~Mbps, Fabric's throughput is similar to the one on Xeon(edge). If the
bandwidth is further limited to 200 and 100~Mbps, the throughput drastically
decreases to around 350 and 170 tps, respectively. This can be explained by the
interplay between transaction size, request rate, and networking bandwidth. In
our experiments, each transaction operates on a record (key-value) of around
1~kB. The endorsement policy is \textit{AND}, meaning that each of the six peers
needs to endorse the transaction and send the read-write sets back to the
client. When the client submits the transaction to the ordering service, the
overhead is $(N+1)\times$ due to the read-write states endorsed by the $N$ peers
plus the original transaction. Let us take the example of the 3,000 tps request
rate. This leads to at least 21 MB/s sent to the ordering service, without
considering the overheads of meta-data. This is close to the 25 MB/s bandwidth
on a 200~Mbps link.

\begin{figure}[tp]
\centering
\includegraphics[width=0.48\textwidth]{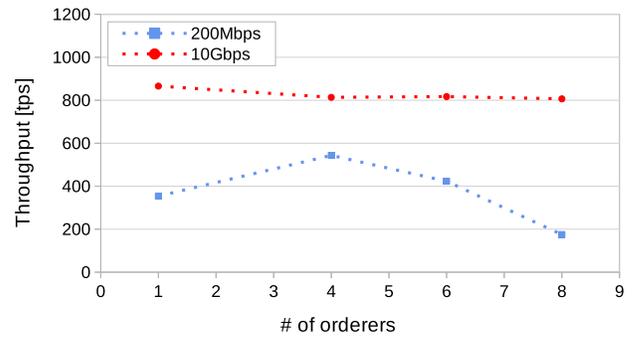}
\caption{Throughput of Fabric with increasing number of orderers.}
\label{fig:fabric-orderers-networking}
\end{figure}

At a closer look, we observe that Fabric's ordering service is a bottleneck
because all the transactions and their endorsements are sent to this service.
But in our default setup, we use a single orderer. Hence, we increase the number
of orders, selecting setups with 4, 6, and 8 orders. We then run experiments in
both the 10 Gbps and 200 Mbps networks. As shown in
Figure~\ref{fig:fabric-orderers-networking}, increasing the number of orderers
to a certain value may help in a low-bandwidth network. For example, having four
orderers instead of one leads to a change in throughput from 350 tps to 550 tps
in a 200 Mbps network. But increasing the number of orders further leads to a
drop in throughput. For example, using 8 orderers in a 200 Mbps network results
in 175 tps. This is because of Raft overhead which grows with the number of
nodes that need to follow the leader. This overhead impacts the performance of
Fabric even in high-bandwidth networks, as shown in
Figure~\ref{fig:fabric-orderers-networking} for the 10 Gbps link.

Similarly, we can explain the drop in throughput shown by 10 Xeon(edge) peers in
Figure~\ref{fig:fabric-tps}. When we limit the bandwidth of Xeon(cloud) to
1~Gbps, we obtain a throughput of 607 tps on 10 peers, similar to 588 tps on
Xeon(edge). This, together with the results presented above, highlight the
trade-off between networking bandwidth and the number of orders in Fabric. In
particular, it is not always possible to have high bandwidth networking in the
real world, especially in cross-continent setups. For example, our measurements
on AWS cloud expose bandwidths of around 10~Mbps across continents. Hence,
running Fabric in a cross-continent setup is challenging.

\begin{figure}[tp]
\centering
\includegraphics[width=0.48\textwidth]{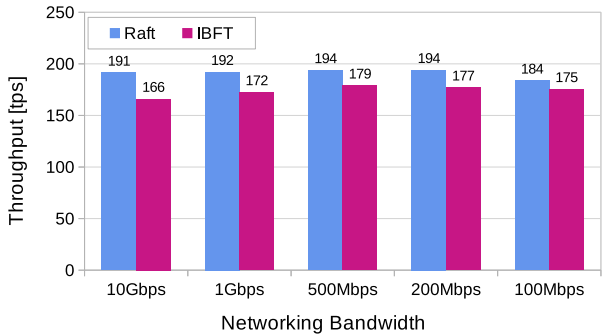}
\caption{Effect of networking bandwidth on Quorum.}
\label{fig:quorum-networking}
\end{figure}

In contrast to Fabric, networking has a negligible impact on the performance of
Quorum, as shown in Figure~\ref{fig:quorum-networking}. Similar to the analysis 
of Fabric, we use Xeon(cloud) and the
\textit{tc} Linux tool to limit the bandwidth of the networking interface on
each Xeon(cloud) peer of the Quorum network. When using Raft consensus, we
observe a small (5\%) drop in throughput only in a 100 Mbps networking link.
When using IBFT, the best performance is achieved in a 500 Mbps link, while in a
10 Gbps network, the throughput of Quorum(IBFT) is 7\% lower compared to the 500
Mbps network. However, these values are within the standard deviation of the
measurements, thus, we do not attribute the throughput drop to networking
bandwidth.

Next, we answer the second part of \textbf{Question~\ref{qst:q2}}, namely, why
does Xeon(edge) have a higher performance than Xeon(cloud) when running Quorum?
Concretely, Xeon(edge) exhibits up to 15\% and 26\% improvement over Xeon(cloud)
when running Quorum(Raft) and Quorum(IBFT), respectively. We attribute this
difference to the higher CPU performance of Xeon(edge) compared to Xeon(cloud).
Recall that, in Section~\ref{sec:sys_char}, we show that Xeon(edge) is 22\%
faster than Xeon(cloud) in ECSDA operations. Since the signing operations  take
the most time in Quorum's execution compared to other functions, as shown in
Listing~\ref{lst:perf_quorum}, Xeon(edge) has an advantage over the other
systems.

Fabric uses a different curve for its ECDSA operations, namely
\textit{secp256r1} as opposed to the \textit{secp256k1} in Quorum.  Our
measurements  in Section~\ref{sec:sys_char} show that it is less optimized. In
particular, Figure~\ref{fig:ecdsa} shows that the IPC of \textit{secp256k1} is
higher compared to \textit{secp256r1}. Nonetheless, the ECDSA operations take
the most time in Fabric's execution compared to other functions, as shown in
Listing~\ref{lst:perf_fabric}.

To answer \textbf{Question~\ref{qst:q3}}, we take a closer look at the
performance of Fabric on TX2. As shown in Figure~\ref{fig:fabric-tps}, TX2
exhibits a throughput that is $6-9\times$ lower compared to Graviton. In
contrast, TX2 exhibits a throughput that is only around $2\times$ smaller
compared to Graviton when running Quorum. For Quorum, the reason is clear: TX2's
CPU core is around $2\times$ slower compared to Graviton's CPU core. But for
Fabric, there is an interplay among all the system's components. As shown
previously, Fabric's execution is memory-intensive and the memory of TX2 has
almost $5\times$ lower bandwidth compared to the memory of Graviton. Hence, when
using 4 and 6 peers, the lower performance of the CPU and memory of TX2 lead to
lower Fabric throughput. On 8 and 10 nodes, besides the CPU and memory, there is
the 1 Gbps networking link of TX2 that hinders the throughput. This is similar
to our analysis on networking bandwidth for Xeon(edge) versus Xeon(cloud). We
also note that a similar analysis applies to RP4, just that this edge device has
even lower performance compared to TX2.
     

\begin{lstlisting}[float=tp,style=customc,frame=single,caption=Example of \texttt{perf report} output for Quorum(IBFT) on Xeon(edge).,label={lst:perf_quorum}]
5.08%  geth [.] secp256k1_fe_sqr_inner                                                                                                                                                         
4.07%  geth [.] runtime.scanobject                                                                                                                                                             
3.99%  geth [.] runtime.findObject                                                                                                                                                             
3.95%  geth [.] secp256k1_fe_mul_inner                                                                                                                                                         
3.89%  geth [.] github.com/ethereum/go-ethereum/vendor/golang.org/x/crypto/sha3.keccakF1600                                                                                                    
2.52%  geth [.] runtime.greyobject                                                                                                                                                             
1.63%  geth [.] github.com/ethereum/go-ethereum/rlp.(*encbuf).encodeString                                                                                                                     
1.55%  geth [.] secp256k1_ge_set_xo_var
...
\end{lstlisting}


\begin{lstlisting}[float=tp,style=customc,frame=single,caption=Example of
\texttt{perf report} output for Fabric on Xeon(edge).,label={lst:perf_fabric}] +  
15.06% peer [.] p256MulInternal                                                                                                                                                                                                                                                                                                                                                     
 8.82% peer [.] p256SqrInternal                                                                                                                                                                                                                                                                                                                                                     
 5.01% peer [.] runtime.mallocgc                                                                                                                                                                           
 2.62% peer [.] runtime.scanobject                                                                                                                                                                         
 2.56% peer [.] github.com/golang/protobuf/proto.unmarshalUint64Value                                                                                                                                      
 2.49% peer [.] crypto/elliptic.p256Select                                                                                                                                                                 
 2.38% peer [.] runtime.adjustframe                                                                                                                                                                        
 2.25% peer [.] runtime.entersyscall                                                                                                                                                                       
 2.19% peer [.] crypto/elliptic.p256PointAddAffineAsm                                                                                                                                                                                                                                                                                                                       
\end{lstlisting}

\begin{figure*}[tp]
	\centering
    \begin{subfigure}{0.48\textwidth}
	\includegraphics[width=0.99\textwidth]{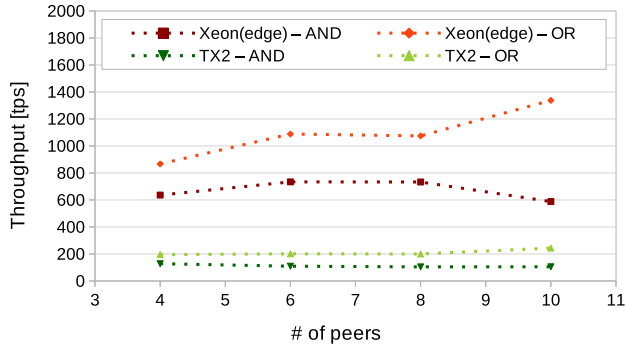}
	\caption{Throughput}
    \label{fig:fabric-policy-tps}
    \end{subfigure}
    ~
    \centering
    \begin{subfigure}{0.48\textwidth}
	\includegraphics[width=0.99\textwidth]{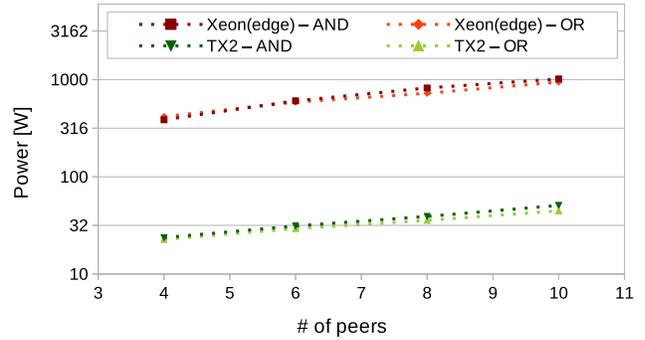}
	\caption{Power}
    \label{fig:fabric-policy-power}
    \end{subfigure}
    \caption{Effect of endorsement policy on Fabric.}
\label{fig:fabric-policy}
\end{figure*}

Lastly, we evaluate the impact of the endorsement policy on the performance of
Fabric. As mentioned before, we use an \textit{AND} policy by default in our
experiments. This means that all the peers in the network need to execute and
endorse a transaction. At the other extreme, there is the \textit{OR} policy
which means that only one peer needs to endorse a transaction. As such, it is
expected that \textit{OR} policy should yield higher throughput due to higher
transaction execution parallelism. Indeed, the throughput of Fabric on 4, 6, 8,
and 10 Xeon(edge) peers, respectively, is 1.4, 1.5, 1.5, and 2.3 times higher
when using \textit{OR} compared to \textit{AND} policy, as shown in
Figure~\ref{fig:fabric-policy-tps}.
In addition, with more peers in the network, the throughput continues to grow as
opposed to dropping in the case of \textit{AND} policy. This can be explained by
both the increasing parallelism in the execution phase and the smaller message
size in the ordering phase which leads to lower networking overhead. Note that
only one endorsement needs to be forwarded to the ordering service when using
\textit{OR} policy.

On the other hand, the improvement in throughput on TX2 when using \textit{OR}
endorsement policy is between $1.5\times$ higher on 4 peers and $2.3\times$
higher on 10 peers. This gap is because (i) the throughput of Fabric with
\textit{OR} policy slowly increases due to higher parallelism, and (ii) the
throughput of Fabric with \textit{AND} policy decreases due to networking
overhead. Hence, the gap becomes bigger.

\subsection{Power Analysis}
\label{sec:power}

In this section, we analyze the power usage of the edge-based systems, namely
Xeon(edge), TX2, and RP4, when running Fabric and Quorum. First, we observe that
Fabric is more power-hungry than Quorum. For example, Xeon(edge) uses around
$1.4\times$ and $1.5\times$ more power to run Fabric compared to Quorum(Raft)
and Quorum(IBFT), respectively. TX2 uses up to $1.6\times$ and $1.8\times$ more
power to run Fabric compared to Quorum(Raft) and Quorum(IBFT), respectively.
Finally, RP4 uses up to $1.2\times$ more power to run Fabric compared to both
Quorum(Raft) and Quorum(IBFT). This is because Fabric is more CPU- and
memory-intensive compared to Quorum. Recall that Fabric uses 1.3 CPU cores, on
average, compared to 0.6 and 0.8 CPU cores used by Quorum(Raft) and Quorum
(IBFT), respectively. Moreover, it is well-known that the CPU power accounts for
the most significant part of a system's power~\cite{GoogleDatacenter_13,
Dumi_PARCO_2019}.

Second, Quorum(Raft) uses more power than Quorum(IBFT), as shown in
Figure~\ref{fig:quorum-power}. In particular, Quorum(Raft) uses up to 13\%,
13\%, and 1\% more power that Quorum(IBFT) on Xeon(edge), TX2, and RP4,
respectively. This is intriguing since Quorum(IBFT) has a slightly higher CPU
utilization compared to Quorum(Raft). But at a closer look using Linux
\textit{perf} tool, we observe that the IPC of Quorum(Raft) is higher compared
to Quorum(IBFT). This suggests that some of the cycles of Quorum(IBFT)'s
execution are spent on waiting for cache/memory operations and these cycles
usually incur less power compared to the execution of other
instructions~\cite{Tudor_2013, Tudor_2014}. Indeed, Quorum(IBFT) exhibits up to
30\% more cache references compared to Quorum(Raft), out of which 21\% are cache
misses.

For Fabric, there is a slight increase in power when \textit{AND} endorsement
policy is used compared to \textit{OR} policy, as shown in
Figure~\ref{fig:fabric-policy-power}. Specifically, there is up to 13\% and 14\%
increase in power when using \textit{AND} policy on Xeon(edge) and TX2,
respectively. But this is expected because \textit{AND} policy uses all the
peers during the endorsement phase, thus, incurring more power compared to
\textit{OR}.

\begin{table}[tp] \centering
\caption{Performance-to-power Ratio of the edge systems.}
\label{table:ppr}
\begin{tabular}{|l|r|r|r|r|r|r|r|r|}
\hline
\multirow{3}{*}{\textbf{Blockchain}} & \multicolumn{2}{c|}{\textbf{Xeon(edge)}}
& \multicolumn{2}{c|}{\textbf{TX2}} & \multicolumn{2}{c|}{\textbf{RP4}} \\
\cline{2-7} & \multirow{2}{*}{\textbf{Peers}} & \textbf{PPR} &
\multirow{2}{*}{\textbf{Peers}} & \textbf{PPR} & \multirow{2}{*}{\textbf{Peers}}
& \textbf{PPR} \\
& & [tpj] & & [tpj] & & [tpj] \\
\hline
\hline
Fabric & 6 & 1.2 & 4 & 5.4 & 4 & 4.3 \\
Quorum(Raft) & 10 & 0.34 & 10 & 2.4 & 10 & 1.1 \\
Quorum(IBFT) & 6 & 0.5 & 6 & 2.4 & 6 & 1.1 \\
\hline
\end{tabular}
\end{table}

Third, we compare the power efficiency across the three edge systems under test.
Power efficiency is reflected by the performance-to-power ratio (PPR), computed
as
\begin{equation} 
PPR=\frac{Throughput}{Power}=\frac{Transactions}{Energy}
\end{equation} 
and expressed in transactions per Joule (ppj). For this, we select the number of
peers that yield the best performance and present the results in
Table~\ref{table:ppr}. We observe that TX2 exhibits the best PPR across all
systems and when running all three blockchains. Compared to Xeon(edge), TX2
exhibits better PPR because it uses much lower power, as shown in
Figure~\ref{fig:fabric-power} and Figure~\ref{fig:quorum-power}.
Compared to RP4, TX2 exhibits better PPR because it yields higher performance
using just a bit more power. For example, TX2 exhibits 35\% higher throughput
while using just 7\% more power when running Fabric on 4 peers compared to RP4.

These PPR results give rise to \textbf{Question~\ref{qst:q4}}, namely, why is
the power efficiency of RP4 lower compared to TX2? The answer to this question
is indicated in our system analysis in Section~\ref{sec:sys_char}. For example,
one CPU core of RP4 has between 16\% and 30\% lower performance compared to one
TX2 CPU core, depending on the benchmark considered. For example, one TX2 core
exhibits 16\% higher performance while using 17\% more energy compared to RP4
when running CoreMark. When running cryptography operations, TX2 has even higher
performance while using slightly higher power. Hence, its PPR is higher compared
to RP4. Part of the lower performance can be attributed to the 26\% lower clock
frequency of RP4. Another source of inefficiency is the memory hierarchy. For
example, the memory bandwidth of TX2 is $3\times$ higher compared to RP4.

\subsection{Cost Analysis}
\label{sec:cost}

In this section, we answer \textbf{Question~\ref{qst:q5}} by conducting a cost
analysis across both edge and cloud systems. This analysis is useful for
permissioned blockchain users that want to host their own nodes. It helps the
users in determining whether to host these nodes at the edge or on the cloud,
and on what type of system. On the cloud, the utilization of an instance is
usually billed every second, based on a price that may vary according to the
datacenter location. At the edge, we use a cost model to estimate the total cost
of ownership (TCO) which considers the fixed hardware cost, the continuing
energy cost, and the manpower cost for system administration. To this end, we
adopt the cost model used in our previous works~\cite{Dumi_BDSN_VLDB15,
Dumi_EDGE_2017}.

\begin{table}[tp] 
\centering
\caption{Cost model parameters.}
\label{table:cost}
\resizebox{0.485\textwidth}{!} {
\begin{tabular}{|l|l|r|r|r|r|}
\hline
\multirow{2}{*}{\textbf{Notation}} &
\multicolumn{2}{c}{\multirow{2}{*}{\textbf{Description}}} &
\multicolumn{3}{|c|}{\textbf{Values}} \\
\cline{4-6} & \multicolumn{2}{c|}{} & \textbf{Xeon(edge)} & \textbf{TX2} &
\textbf{RP4} \\
\hline
\hline
$T$ & \multicolumn{2}{l|}{cluster lifetime} & \multicolumn{3}{c|}{3 years
(26,280 hours)} \\
\cline{4-6} $N$ & \multicolumn{2}{l|}{number of cluster nodes} &
\multicolumn{3}{c|}{10 (6)} \\
$C_{ph}$ & \multicolumn{2}{l|}{cost of electricity per hour [USD]} &
\multicolumn{3}{c|}{0.10} \\
$C_{mh}$ & \multicolumn{2}{l|}{cost of manpower per hour [USD]} &
\multicolumn{3}{c|}{6.85 (5,000 per month)}
\\
\hline
$C_{s}$ & \multicolumn{2}{l|}{cost of acquisition per node [USD]} & 5,000 & 800
& 100 \\
\hline
\multirow{3}{*}{$P_{a}$} & average power & Fabric & 101.6 & 5.1 & 4.2 \\
& \multicolumn{1}{c|}{per node} & Quorum(Raft) & 73.4 & 4.1 & 2.4 \\
& \multicolumn{1}{c|}{[W]} & Quorum(IBFT) & 65 & 3.2 & 3.7 \\
\hline
\end{tabular}
}
\end{table}

The edge TCO model adopted in this paper consists of the equipment cost over a
period of time, which is usually three years~\cite{Dumi_BDSN_VLDB15}, and the
cost of electricity,
\begin{equation}
C = N \cdot C_s + N \cdot T \cdot P_a \cdot C_{ph} + T \cdot C_{mh}
\end{equation}
where $N$ is the number of nodes, $C_s$ is the cost of buying one node, $T$ is
the lifespan of the cluster in hours, $P_a$ is the average power of one node in
kilo-Watts (kW), $C_{ph}$ is the electricity cost per kilo-Watt-hour (kWh),
$C_{mh}$ is the cost of manpower per hour. The assumption is that node
utilization, hence, the average power is constant throughout the lifespan of the
cluster. In reality, the power depends on the request rate, but here we suppose
we have a high enough request rate to keep the blockchain busy. To compare edge
nodes with cloud nodes, we divide the TCO by the number of nodes and the
lifespan in hours to get the average price per hour for a single edge node.

\begin{figure}[tp]
\centering
\includegraphics[width=0.48\textwidth]{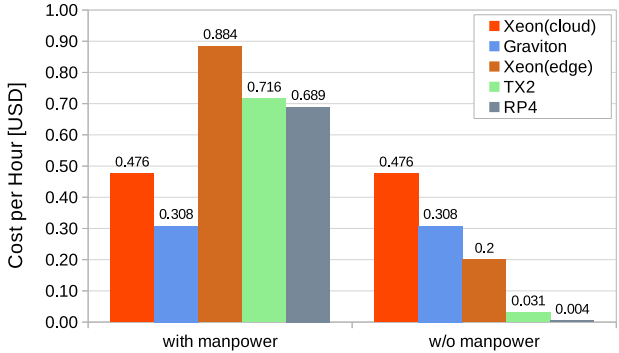}
\caption{Comparison of cost across all systems.}
\label{fig:cost-per-hour}
\end{figure}

The model parameters are summarized in Table~\ref{table:cost}. For a fair
comparison with the cloud costs used in this paper, which represent the AWS
region hosted in Oregon, we use the price of electricity in Oregon, which is
\$0.09 USD for commercial entities~\cite{oregon_energy}. For manpower cost, we
use \$5,000 per month, which is in the lower range of the salary of an Amazon
datacenter technician~\cite{aws_salary}. The acquisition price of the nodes is
based on our own records. The number of nodes and the corresponding average
power per node depend on the blockchain framework. We select the number of nodes
that yield the best performance, which is 10 for Fabric and Quorum(Raft), and 6
for Quorum (IBFT).

When computing the TCO, we consider two scenarios. First, we assume that there
needs to be a technician in charge of setting up and providing maintenance for
the hardware nodes. Second, we assume that both the edge and the cloud need a
system administrator to set up and operate the nodes, hence, we do not include
this manpower cost in the TCO. Here, we assume that the cloud is used as
Infrastructure-as-a-service (IaaS). The price per hour per node for Fabric is
shown in Figure~\ref{fig:cost-per-hour} for both these scenarios. Our key
observation is that the edge is much cheaper than the cloud when manpower cost
is disregarded. For example, Xeon(edge), TX2, and RP4 are respectively
$1.5\times$, $10\times$, and $75\times$ cheaper than Graviton. On the other
hand, adding manpower cost results in almost $2\times$ higher cost of the edge
nodes compared to the cloud nodes. For example, Xeon(edge) costs \$0.884 per
hour, while Xeon(cloud) costs \$0.476 per hour. These results show that the cost
of continuous operation due to energy has little impact in the long run.
Instead, the hardware and manpower costs have a high impact, and this may lead
the users to choose cloud computing over on-premise setups.

\section{Conclusions}
\label{sec:concl}

In this paper, we conducted an in-depth performance evaluation of two
widely-used permissioned blockchains on a diverse range of hardware systems. The
two selected blockchains, Hyperledger Fabric and ConsenSys Quorum represent
execute-order-validate and order-execute transaction flows, respectively. In
addition, we analyze the performance of both Quorum with Raft and Quorum with
IBFT to represent both CFT and BFT application scenarios. The hardware systems
represent both the cloud and the edge, as well as, both x86/64 and ARM
architectures. For example, we use cloud instances with Intel Xeon and ARM-based
Graviton CPUs, edge servers with Intel Xeon CPU, and edge devices such as Jetson
TX2 and Raspberry Pi 4.

Among the key observations, we show that ARM-based Graviton cloud instances
achieve higher performance than Xeon-based instances due to impressive CPU and
memory performance. Moreover, these instances are around 35\% cheaper than their
Xeon counterparts. On the other hand, edge devices based on ARM CPUs exhibit
much lower performance but higher power efficiency compared to Xeon-based
servers. Depending on the application scenario, if high throughput is not
required, then low-power edge nodes can be used to run the blockchain. In the
long run, hosting blockchain nodes at the edge (on-premise) may lead to
substantial savings in cost, if the manpower cost is not included in the total
cost of ownership.

\section*{Acknowledgments}

This research is supported by the National Research Foundation, Singapore under
its Emerging Areas Research Projects (EARP) Funding Initiative. Any opinions,
findings and conclusions or recommendations expressed in this material are those
of the author(s) and do not reflect the views of National Research Foundation,
Singapore.

\balance



\end{document}